\documentclass[12pt]{iopart}

\usepackage{graphicx}
\usepackage{cite}
\usepackage{color}
\usepackage{multirow}
\usepackage[section]{placeins}
\usepackage{amssymb}

\begin{document}

\setlength{\parindent}{25pt}

\title[Resonance effects of electron transport]{Monte Carlo simulation of resonance effects of electron transport in a spatially modulated electric field in Ar, N$_2$ and their mixtures}

\author{A. Albert$^{1}$, D. Bo\v{s}njakovi\'c$^{2}$, S. Dujko$^{2}$, Z. Donk\'o$^{1}$}

\address{$^1$Institute for Solid State Physics and Optics, Wigner Research Centre for Physics, 1121 Budapest, Konkoly Thege Mikl\'os str. 29-33, Hungary\\}

\address{$^2$Institute of Physics, University of Belgrade, Pregrevica 118, 11080 Belgrade, Serbia}

\ead{donko.zoltan@wigner.hu}

\begin{abstract}
The relaxation of the distribution function of the electrons drifting under the influence of a homogeneous electric field in noble gases is known to take place over an extended spatial domain at `intermediate' values of the reduced electric field, $E/N$. We investigate the transport of electrons in Ar and N$_2$ gases, as well as in their mixtures at such $E/N$ values ($\sim$ 10--40 Td). After discussing briefly the basic scenario of relaxation in a homogeneous electric field, the major part of work concentrates on the properties of transport in an electric field that is spatially modulated within a finite region that obeys periodic boundaries. The spatial distribution of the mean velocity, the mean energy, and the density of the electrons, the importance of the excitation channels, as well as the electron energy distribution function are obtained from Monte Carlo simulations for various lengths of the computational domain, at different mean values and degrees of modulation of the reduced electric field. At low modulations, the spatial profiles of the mean velocity and mean energy are nearly harmonic, however their phases with respect to the electric field perturbation exhibit a complex behaviour as a function of the parameters. With increasing modulation, an increasing higher harmonic content of these profiles is observed and at high modulations where an electric field reversal occurs, we observe trapping of a significant population of the electrons. The effect of mixing a molecular gas, N$_2$, to Ar on the transport characteristics is also examined. Transition to local transport at high N$_2$ admixture concentrations and long spatial domains is observed.
\end{abstract}

\maketitle

\section{Introduction}

Under {\it hydrodynamic} conditions, the velocity distribution function (VDF) of electrons subjected to a homogeneous and stationary electric field is a unique function of the reduced electric field, $E/N$. Up to moderate values of $E/N$ (typically up to few hundred Townsends, 1 Td = 10$^{-21}$\,V\,m$^2$) the VDF of electrons in noble gases generally exhibits only a small anisotropy, i.e. it can be well approximated by an isotropic part and small anisotropic part. This behaviour, which is the basis of the two-term approximation, stems from the effect of frequent elastic collisions of the electrons with the gas atoms that randomize the directions of velocities and thus their characteristic thermal velocity strongly exceeds the average (directed) velocity. In the hydrodynamic regime, the energy gain and loss of the electrons compensate each other exactly, however, in different ways depending on the strength of the reduced electric field \cite{Winkler2002,KumarSR1980}. 

Considering noble gases, at very low $E/N$ ($\lesssim$ 10 Td) the energy loss of electrons is mostly due to elastic collisions. In such events, a fraction of their energy proportional to the electron/atom mass ratio ($m/M$) is lost. For an electron energy of 1 eV, e.g., such a collision results in a loss of $\approx$ 10$^{-5}$ eV, in the case of argon gas. At somewhat higher $E/N$ values ($\sim$tens of Td-s), where the energy of the electrons reaches the threshold for inelastic processes ($\sim$ 10 eV), the main channels of the energy loss become the excitation processes. At such conditions the lowest excited levels can be reached first. With increasing $E/N$, the number of these channels increases and at some point ionization becomes possible, too. Above $\sim$ 100 Td the latter process is usually appreciable.

The spatial evolution of the electron VDF in {\it non-hydrodynamic regimes} is also remarkably different for the $E/N$ rages distinguished above \cite{noneq1,noneq2,noneq3,noneq4}. Non-hydrodynamic transport establishes under various conditions: (i) when the electric field varies over a characteristic length that is short compared to the mean free path of the electrons, (ii) when the temporal change of the field is quicker than the mean time between electron-neutral collisions, (iii) in the presence of sources/sinks of electrons and/or (iv) in the presence of boundaries. In the latter case, even in the presence of a homogeneous and stationary electric field, the VDF and the transport parameters (mean velocity, mean energy, etc.) of the electrons vary in space. A classical example of this scenario is an electrode that emits electrons with a certain initial velocity distribution that is defined by the emission process (e.g. photoemission). This VDF is clearly different from that acquired by the electrons under hydrodynamic conditions for the given $E/N$ that is present in the volume considered. This implies that a transition region ("equilibration region") must exist within which the VDF transforms from its initial shape to the equilibrium shape, see e.g \cite{equi,Donko2011}. 

The fundamental experiment of Franck and Hertz \cite{FH}, which provided evidence for the existence of quantised energy levels of the atoms, actually utilised this effect. Specifically, this experiment focused on the early phase of the equilibration, where a prominent periodic structure in the mean energy of the electrons was present. The experiment was operated in the "window" of $E/N$ values where such behavior prevails. The electron kinetics in this experiment has been investigated in a number of works, e.g. \cite{RobsonLW2000,Sigeneger2003,White2012,Robson2014}. 

We note that both at low and high $E/N$ values no, or less prominent periodic structures can be observed, respectively, due to the smooth transition mediated by elastic collisions, and due to the rapid randomisation of the electron energy in the presence of a high number of inelastic energy loss channels and the possibility of ionization that creates additional particles. The extended spatial structures formed at intermediate $E/N$ values have attracted much attention \cite{Loffhagen2002,Dujko2008,White2009}. The equilibration of an electron swarm in argon gas at $E/N$ values at few tens of Td-s was as well observed experimentally recently in a scanning drift tube apparatus \cite{equilibration2019} that makes it possible to follow the spatio-temporal development of electron swarms. 

Under the conditions, where the electric field is spatially modulated a strong modulation of the electron transport characteristics appears at some $E/N$ values as revealed in studies based on the solution of the Boltzmann equation by Golubovsky {\it et al.} \cite{Golubovsky1998,Golubovskii1999}. In the presence of appreciable charge density, the spatial variation of the transport characteristics can itself give rise to a perturbation of the electric field. As this interplay may be self-amplifying, stationary or moving spatial structures can show up in discharge plasmas. Such structures, often termed as "striations" have thoroughly been investigated for several decades, see, e.g., the review by Kolobov \cite{Kolobov2006}. The early studies based on analytic approaches \cite{Klarfeld1952,Pekarek1962} have later been replaced by kinetic treatment of the electrons \cite{Sigeneger2000a,Sukhinin2006}. Striations, caused by different mechanisms, are present in a variety of plasma sources, like dc glow discharges \cite{Raizer1997}, plasma display panels \cite{Iza2005}, and inductively coupled radiofrequency discharges \cite{Stittsworth1996}. Despite the extensive work done in this field \cite{Sigeneger1998,Sigeneger2000b}, the complex dynamics of striations is still subject of intensive current research, e.g. 
\cite{Liu2016,Arslanbekov2019,Kolobov2020,Harti2020}.

Most of the investigations of the electron kinetics have been based on the solution of the Boltzmann equation \cite{Winkler2002,Golubovskii12013}, particle based simulations were used only in a fewer number of cases. As examples for the latter, studies of  striations in inductively coupled \cite{Denpoh} and capacitively coupled electronegative \cite{Liu2019} plasmas, and in ionization waves in barrier discharges \cite{Shvydky} may be mentioned. Due to the rapid development of computing hardware such particle based methods became equally suited  as the numerical solutions of the Boltzmann equation, for studies of particle transport in spatially varying fields due to their ability to capture fully the nonlocal kinetic effects appearing in various settings. In this paper, we use Monte Carlo simulation (e.g. \cite{Tagashira86,Boeuf82}) to investigate certain aspects of the transport of electrons in spatially varying electric fields. 

The simulation method is discussed in section \ref{sec:sim}. In section \ref{sec:param}, we briefly introduce some important physical quantities, the characteristic momentum and energy relaxation frequencies and lengths, that help understanding the relaxation and resonance effects to be discussed later on. The presentation of the results in section \ref{sec:results} starts with illustrating the spatial relaxation of electron swarms in Ar and in Ar-N$_2$ mixtures in a homogeneous electric field. These findings aid choosing the proper parameter range of the reduced electric field for which the studies of the transport in a periodically modulated electric field are conducted. The results of these simulations for Ar are presented in section \ref{sec:results-ar}. We analyse the spatial profiles of the mean electron energy, velocity and density for various values of the average $E/N$ and illustrate the effect of the modulation of $E/N$ on these profiles. Additionally, we investigate the spatial distribution of the electron energy distribution function and the distribution of the excitation channels (reaction rates). Fourier analysis of the profiles of the mean velocity and energy (i) reveals the phase between them and the modulated field (at low modulation) and (ii) shows how their harmonic content increases at high modulation. In section \ref{sec:mix} the effect of an N$_2$ admixture on the electron transport characteristics is addressed. A brief summary is given in section \ref{sec:summary}.  

\section{Simulation method}

\label{sec:sim}

Our studies are based on the Monte Carlo (MC) description of the motion of electrons in (i) homogeneous and (ii) spatially modulated electric field ${\bf E}(x)$. We use the well-established MC algorithm for charged particle transport (e.g. \cite{Donko2011}) and solve the discretised version of the equation of the motion of the electrons,
\begin{equation}
    m \ddot{\bf r} = q {\bf E}(x),
\end{equation}
using the Velocity-Verlet method, with constant time steps of $\Delta t = $1 ps. Here, $m$ and $q$ are the mass and the charge of the electrons. The probability of a collision to occur during the $\Delta t$ time step is 
\begin{equation}
    P_{\rm coll} = 1 - \exp \bigl[-N \sigma_{\rm tot}(v) v \Delta t \bigr],
\end{equation}
where the total cross section $\sigma_{\rm tot}$ is the sum of the cross sections of all possible collision processes. Whenever a collision occurs, its type is chosen randomly, taking into account the values of all cross sections at the actual velocity of the colliding electron.

The electron - Ar atom cross section set is based on \cite{Hayashi}, includes the elastic momentum transfer cross section, excitation to 25 distinct Ar levels, and the ionisation cross section. The cross section set for electron - N$_2$ molecule collisions are taken from \cite{Biagi}. The set includes the elastic momentum transfer cross section, excitation to several vibrational and electronic states of N$_2$, as well as  the ionisation cross section. As calculations are performed for $E/N \gg 1$ Td, the cross sections for rotational excitations are not included in the present study. Ionisation is treated here as a conservative process, i.e. just like an excitation event, to ensure that the number of electrons does not grow in the simulations. This approach is justified at the $E/N$ values considered here, where ionisation has a very small rate (which is confirmed by the results). This simplification could easily be omitted when necessary, e.g. at higher $E/N$ values. All collisions are assumed to result in isotropic scattering, the thermal motion of the background gas atoms is disregarded (i.e., the "cold gas approximation" is adopted). The electrons do not interact with each other, i.e. we study classical swarm conditions at low charged particle density. The simulations are conducted at a pressure of $p =$ 100 Pa and at the ambient temperature of $T_{\rm g}$ = 300 K, i.e. at a neutral density of $N \cong 2.42 \times 10^{16}$ cm$^{-3}$. 

\begin{figure}[ht ]
\begin{center}
\includegraphics[width =0.5\textwidth]{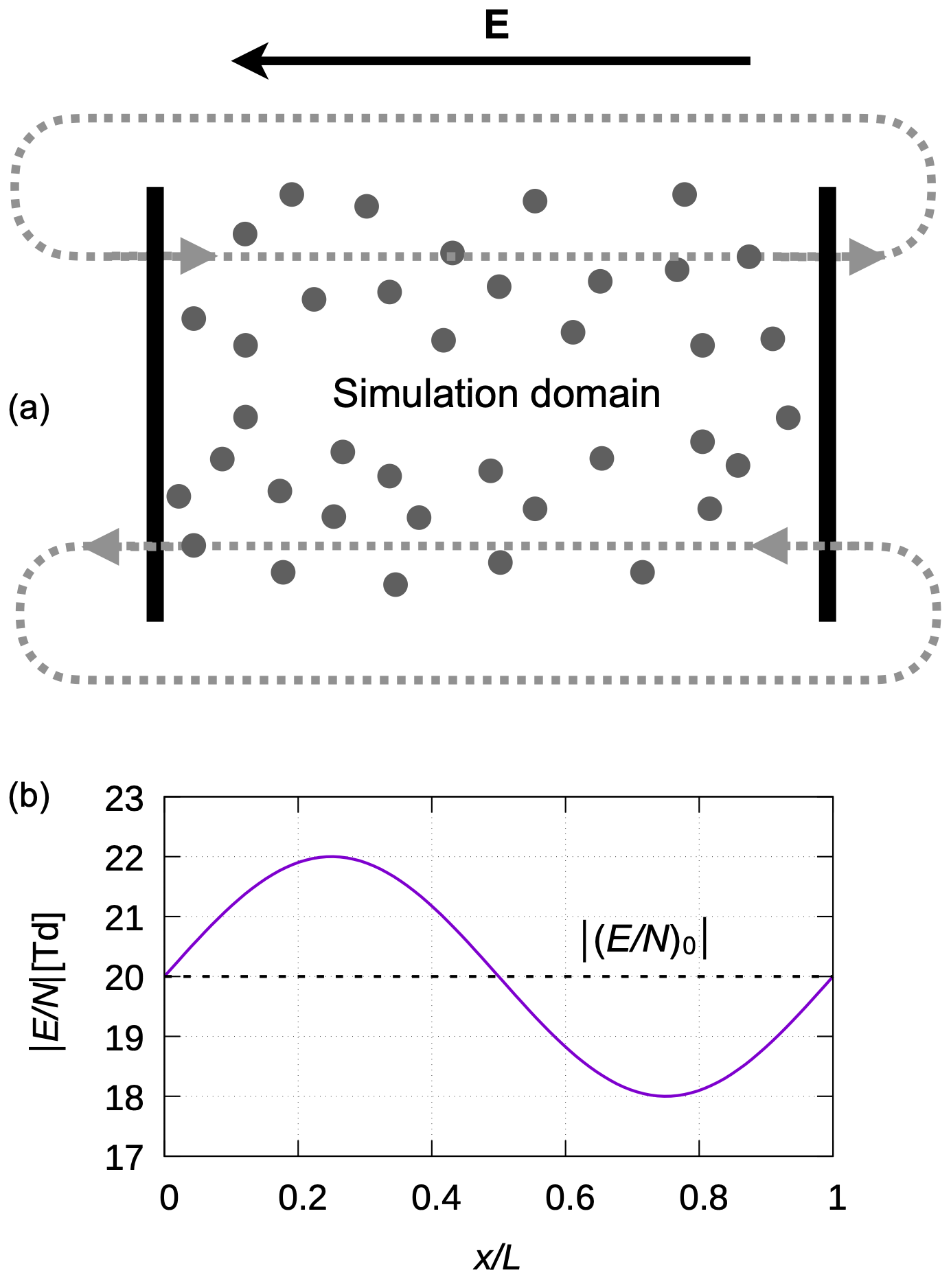}
\caption{(a) In the case of a modulated electric field, the electrons' motion is followed in a simulation box having a width $L$, with periodic boundaries. The system is exposed to a sinusoidally spatially modulated electric field as shown in (b) for the case of an average reduced electric field of $(E/N)_0$ = 20 Td (indicated by the dashed horizontal line) and a modulation depth $M$ = 0.1.}
\label{fig:simsetup}
\end{center}
\end{figure}

Except for the study of swarm relaxation in a homogeneous electric field, the particles are restricted to move within a simulation domain that obeys periodic boundary conditions, as shown in figure \ref{fig:simsetup}(a). Particles leaving this domain in the $\pm x$ directions are re-injected into the domain at the opposite sides. The periodic boundaries emulate an infinite system with spatially periodic modulation of $E/N$.

The electric field points in the $-x$ direction, consequently, the electrons drift in the $+x$ direction. In the following, we omit the negative sign of $E$. In the case of a modulated electric field, the form
\begin{equation}
    E(x) = E_0 \bigl[1 + M\,\sin(k x) \bigr],
    \label{eq:field}
\end{equation}
is adopted, where $E_0$ is spatial average of the electric field, $M \geq 0$ is the modulation depth, and $k = 2\pi / L$, with $L$ being the length of the spatial period of modulation (see figure \ref{fig:simsetup}(b)). The voltage drop over the length $L$ is $U = (E/N)_0 \,N\, L$, i.e.
\begin{equation}
    U {\rm [V]} = 0.242 \cdot (E/N)_0 {\rm [Td]} \cdot L {\rm [cm]}.
    \label{eq:voltage}
\end{equation}
At a given $(E/N)_0$ this relation connects the length of the computational domain and the voltage drop over this domain. We primarily examine the range of the parameters $(E/N)_0$ and $L$, where the energy corresponding to the voltage drop over the simulation domain, $q\,U$, is in the order of the excitation levels of Ar, the lowest being 11.55 eV. 

We present results for the spatial profiles of the mean electron velocity $\overline{v}(x)$, the mean energy $\overline{\varepsilon}(x)$, the electron density $n(x)$, the electron energy distribution function (EEDF) $f(x,\varepsilon)$. These characteristics are studied as a function of the reduced electric field and the length of the computation cell (that equals the wavelength of the modulation of $E(x)$.) They are "measured" within 200 slabs with equal width, covering the simulation domain of length $L$. A lower spatial resolution is used for the analysis of the excitation rates and of the spatial distribution of the EEDF, to ensure a better statistics. 

At low values of modulation, the system is expected to give a linear response for the space-dependent scalar quantities, i.e. the latter are foreseen to exhibit a harmonic spatial profile. With increasing perturbation, a nonlinear response is expected to establish. Taking as an example the mean velocity $\overline{v}(x)$, the harmonic content as well as the phase shift of the individual harmonics contributing the spatial profile can be obtained via Fourier analysis, which allows to construct $\overline{v}(x)$ as
\begin{equation}
   \overline{v}(x) = \sum_s \overline{v}_s \sin (s\,k\, x -\varphi_s),
\end{equation}
where $\overline{v}_s$ and $\varphi_s$ are, respectively, the amplitude and the phase delay of the $s$-th harmonic. This analysis helps, e.g., identifying the conditions when local transport is approached: in the case of low modulation (as long as $\overline{v}$ and $\overline{\varepsilon}$ are monotonically increasing functions of $E/N$) we expect $\overline{v}_s \rightarrow 0, \forall s>1$ and $\varphi_1 \rightarrow 0$,  as $E(x)$ contains only a dc component and one (perturbing) harmonic with $s=1$. Any deviation from this behaviour is the signature of the non-local character of the transport and the nonlinear response of the system to the electric field perturbation. We note that this analysis of the phase shifts of the "macroscopic quantities", like $\overline{v}(x)$ and $\overline{\varepsilon}(x)$ with respect to $E(x)$ does not offer an explanation for the resonance effects, as these are kinetic by nature. We also need to notice that the dependence of $\overline{v}$ on $E/N$ is not necessarily monotonic, this scenario is called Negative Differential Conductivity (NDC) \cite{NDC}, which typically occurs in gas mixtures, including Ar-N$_2$ mixtures that are also studied here \cite{Dyatko}.


\section{Relaxation frequencies and lengths}

\label{sec:param}

\begin{figure}[ht ]
\begin{center}
\footnotesize(a) \includegraphics[width =0.45\textwidth]{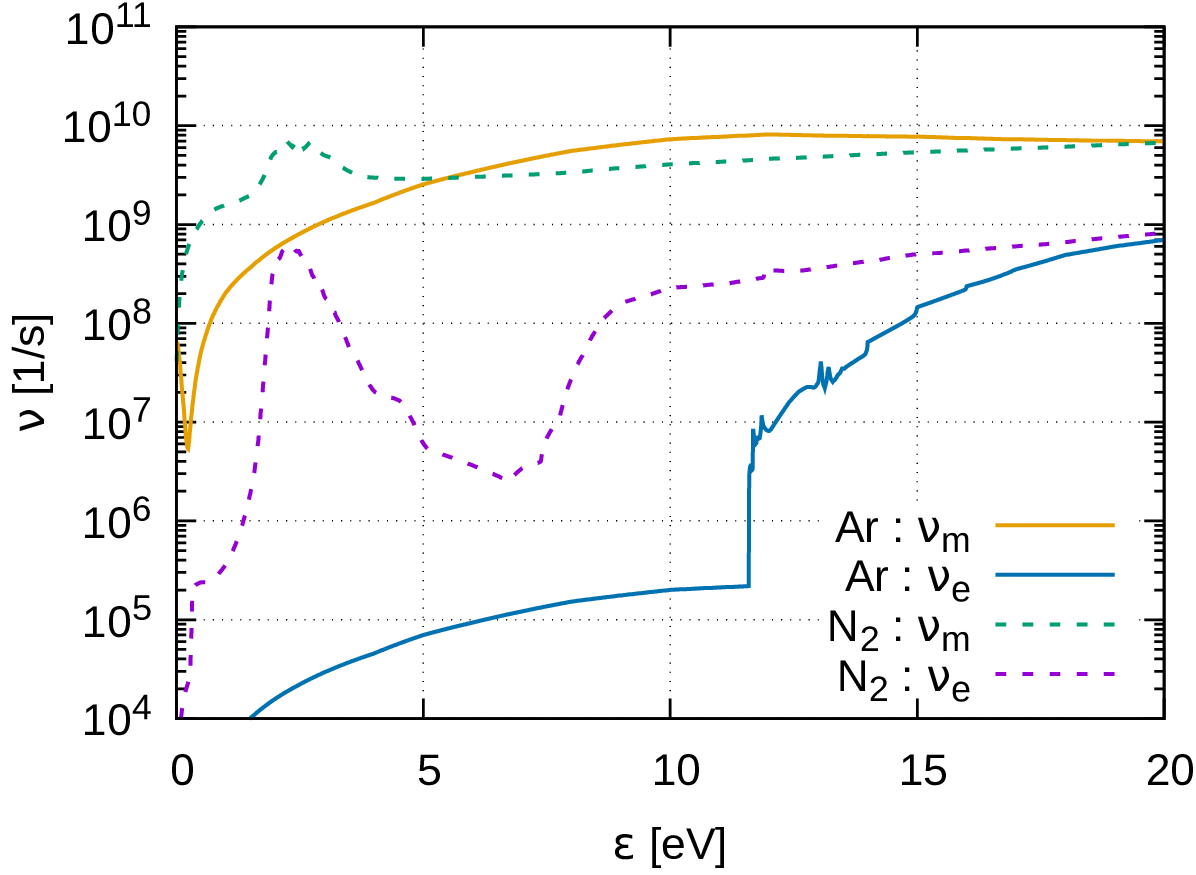}\\
\footnotesize(b) \includegraphics[width =0.45\textwidth]{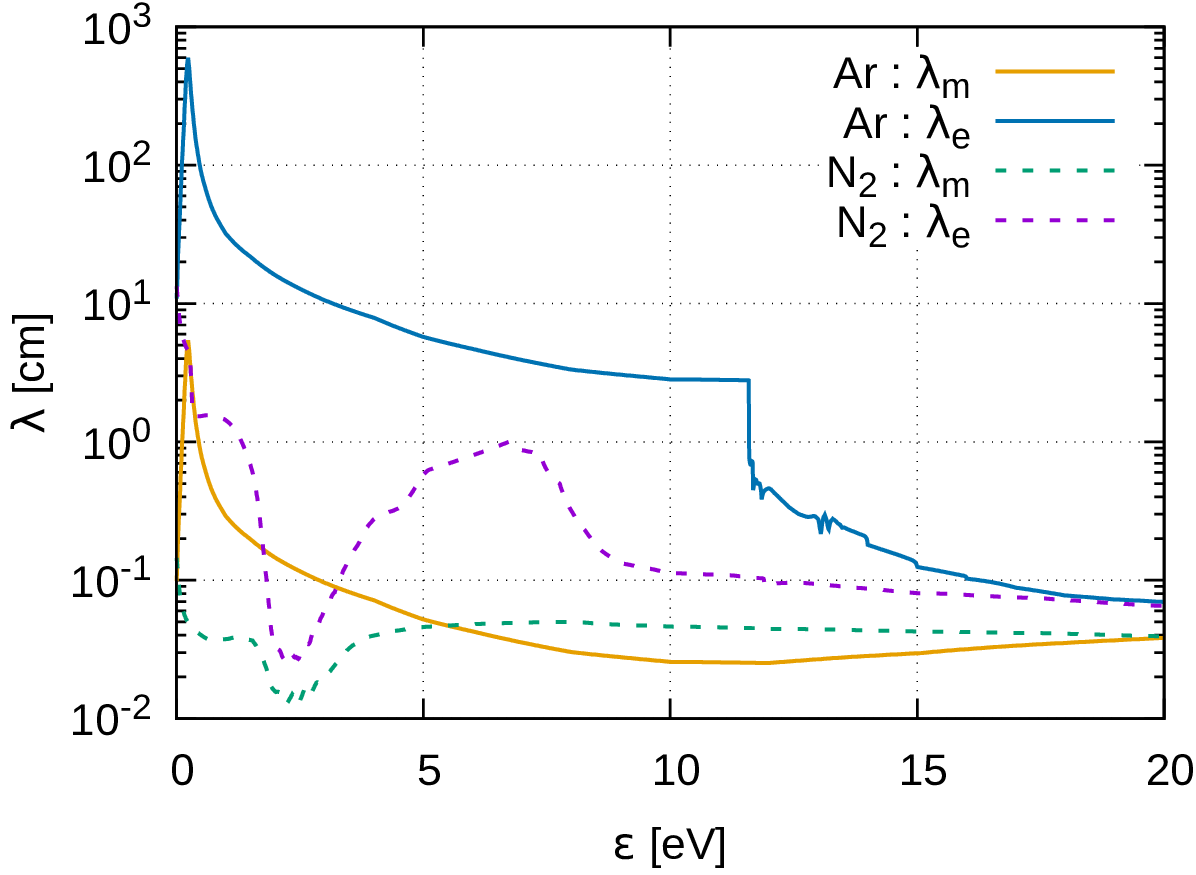}\\
\footnotesize(c) \includegraphics[width =0.45\textwidth]{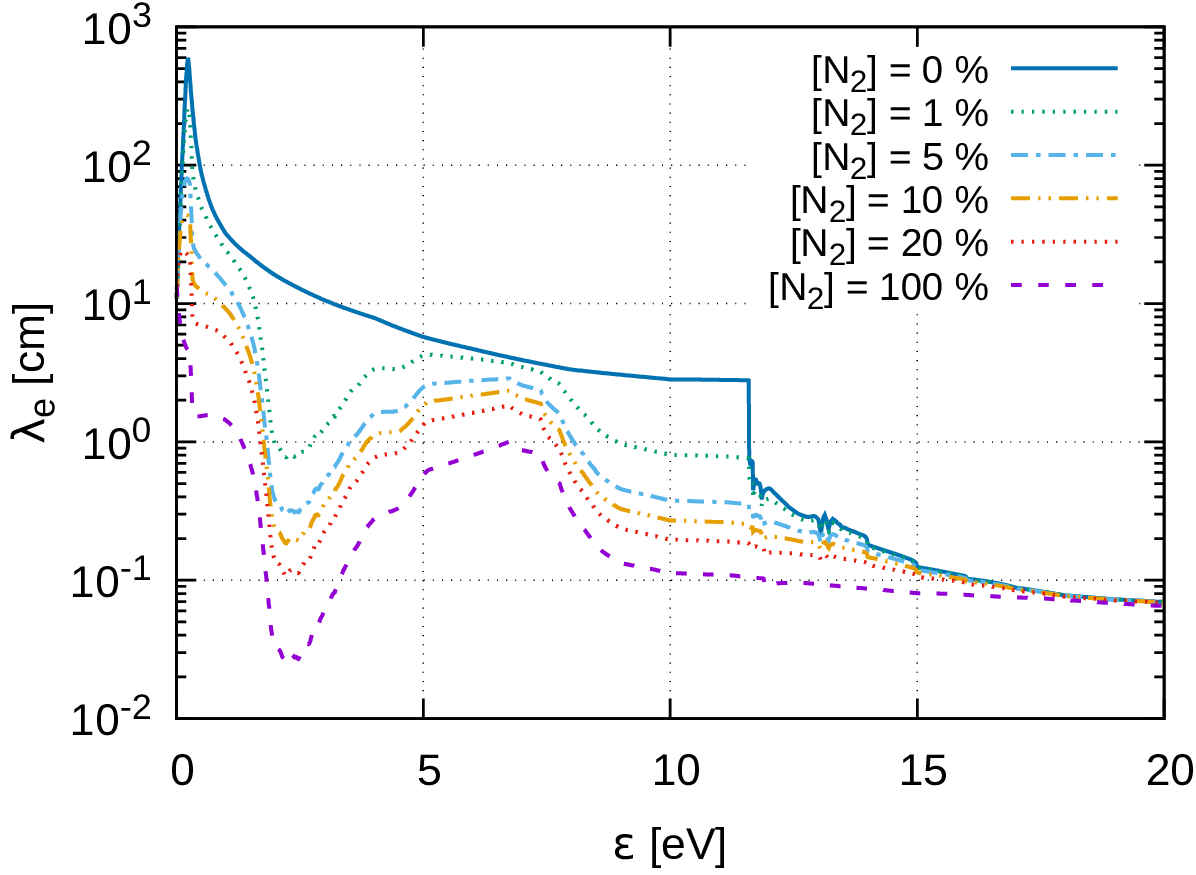}
\caption{(a) Momentum and energy dissipation frequencies $\nu_{\rm m}$ and $\nu_{\rm e}$, as well as (b) mean free path $\lambda_{\rm m}$ and the energy relaxation length $\lambda_{\rm e}$ in Ar (solid lines) and N$_2$ (dashed lines). (c) Dependence of the energy relaxation length on the concentration of N$_2$ in the Ar+N$_2$ mixture. $p$ = 100 Pa and $T_{\rm g} = 300$ K.}
\label{fig:LF}
\end{center}
\end{figure}

Before presenting our results it is useful to illustrate the behaviour of few important quantities that have major influence on the relaxation and resonance effects to be discussed. These are the momentum and energy dissipation frequencies, $\nu_{\rm m}$ and $\nu_{\rm e}$, respectively, as well as the mean free path $\lambda_{\rm m}$ and the energy relaxation length $\lambda_{\rm e}$. We have computed these quantities according to the expressions given in \cite{Winkler2002} and display them in figure \ref{fig:LF}. Panel (a) shows the energy dissipation frequency ($\nu_{\rm e}$) and the momentum dissipation frequency ($\nu_{\rm e}$) for both Ar and N$_2$. We find $\nu_{\rm m}$ to be significantly higher than $\nu_{\rm e}$ over the whole range of energies considered. $\nu_{\rm e}$ is especially low for Ar below the threshold energy for inelastic loss channels, as in elastic collisions the fractional energy loss of the electrons is in the order of the electron/atom mass ratio, as already mentioned in section 1. Rapid changes of $\nu_{\rm e}$ with $\varepsilon$ can be observed for both gases. Whenever the energy distribution of the electrons spans a range that includes such a change, parts of the electron population with different energies will behave dissimilar in terms of energy relaxation, as explained in \cite{Winkler2002}. 

The momentum relaxation frequency does not exhibit abrupt changes as a function of the energy, except for Ar at low energies, due to the Ramsauer-Townsend minimum in the elastic collision cross section. Regarding the relaxation lengths in pure gases, figure \ref{fig:LF}(b) reveals that the energy relaxation length ($\lambda_{\rm e}$) exceeds considerably the mean free path ($\lambda_{\rm m}$). Below the inelastic excitation threshold in Ar, e.g., their ratio amounts about two orders of magnitude. Under such conditions the relaxation of the energy in a swarm is expected to take place over an extended spatial scale, where a high number of collisions is required for equilibration. The difference between $\lambda_{\rm e}$ and $\lambda_{\rm m}$ for Ar decreases as the energy is increased, at 20\,eV the ratio between them drops to a factor of two. As around this energy the sum of inelastic cross sections approaches the value of the elastic momentum transfer cross section energy relaxation becomes efficient. As to N$_2$, $\lambda_{\rm e}$ and $\lambda_{\rm m}$ are relatively close to each other, meaning that energy relaxation takes place over a few free flight lengths of the electrons, except at low energies (below 2\,eV) and within the 3\,eV$\leq \varepsilon \leq$ 8 \,eV interval, where the collision cross sections are low. 

Due to the large disparity of the relaxation frequencies and lengths in Ar vs. in N$_2$, even a small amount of the latter causes a significant change of these parameters, as illustrated in figure \ref{fig:LF}(c) for the case of $\lambda_{\rm e}$. With respect to the case of pure Ar the strongest decrease of $\lambda_{\rm e}$ occurs in the 2\,eV$\leq \varepsilon \leq$ 3 \,eV  and 8\,eV$\leq \varepsilon \leq$ 11 \,eV domains of the electron energy as a result of the addition of N$_2$ to Ar. This is caused by, respectively, the vibrational and electronic excitation of N$_2$ molecules. In the first domain, even 1\% of N$_2$ decreases $\lambda_{\rm e}$ by a factor of 10 as it can be seen in figure \ref{fig:LF}(c). At electron energies above $\approx$ 15\,eV the effect of N$_2$ on $\lambda_{\rm e}$ becomes negligible due to the availability of a high number of inelastic loss channels.

\section{Results}

\label{sec:results}

\subsection{Homogeneous electric field}

\label{sec:reults-homog}

\begin{figure}[ht ]
\begin{center}
\footnotesize(a) \includegraphics[width =0.45\textwidth]{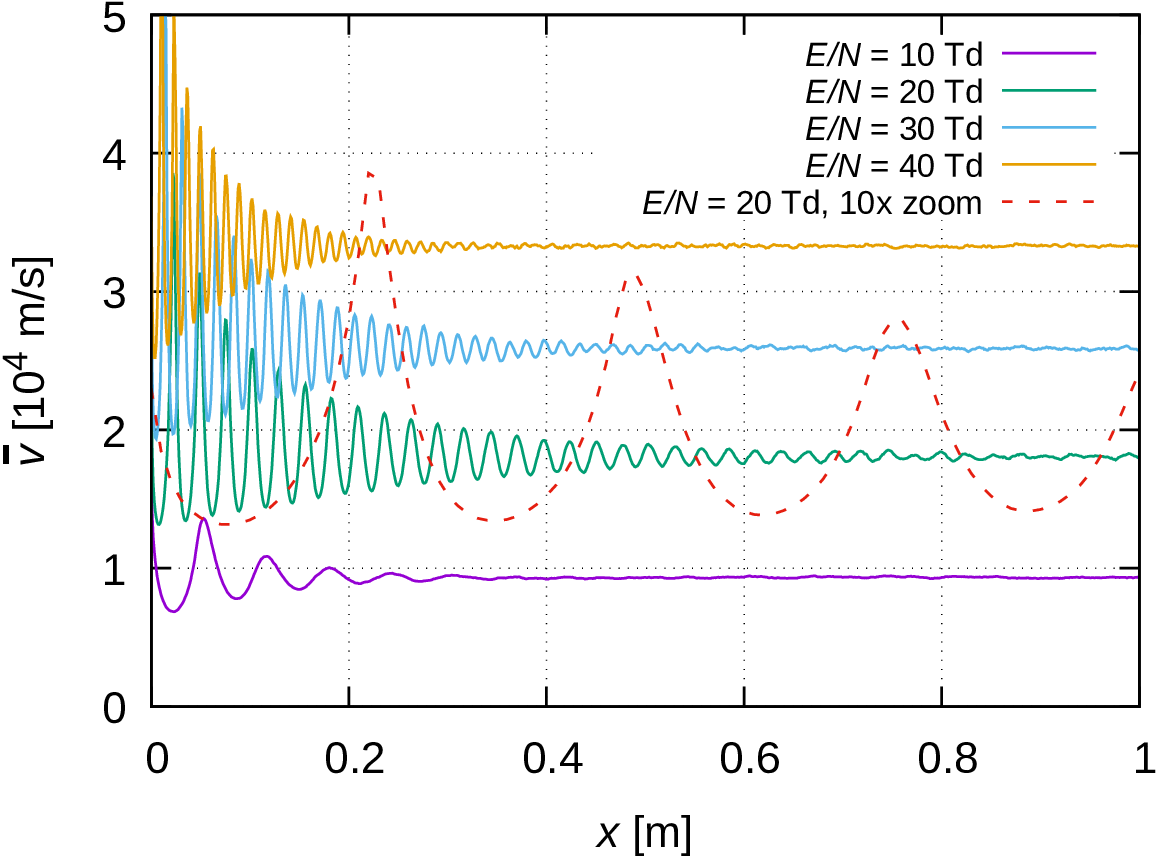}
\footnotesize(b) \includegraphics[width =0.45\textwidth]{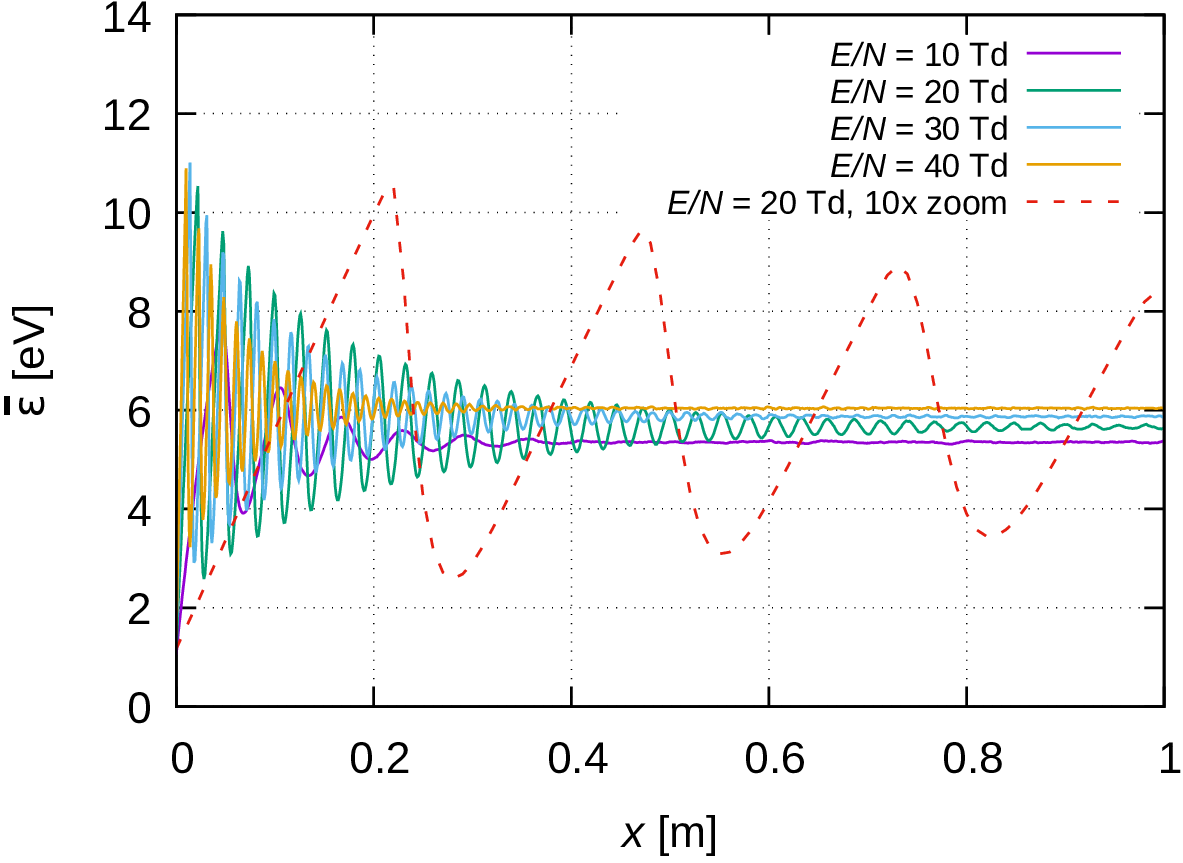}\\
\footnotesize(c) \includegraphics[width =0.45\textwidth]{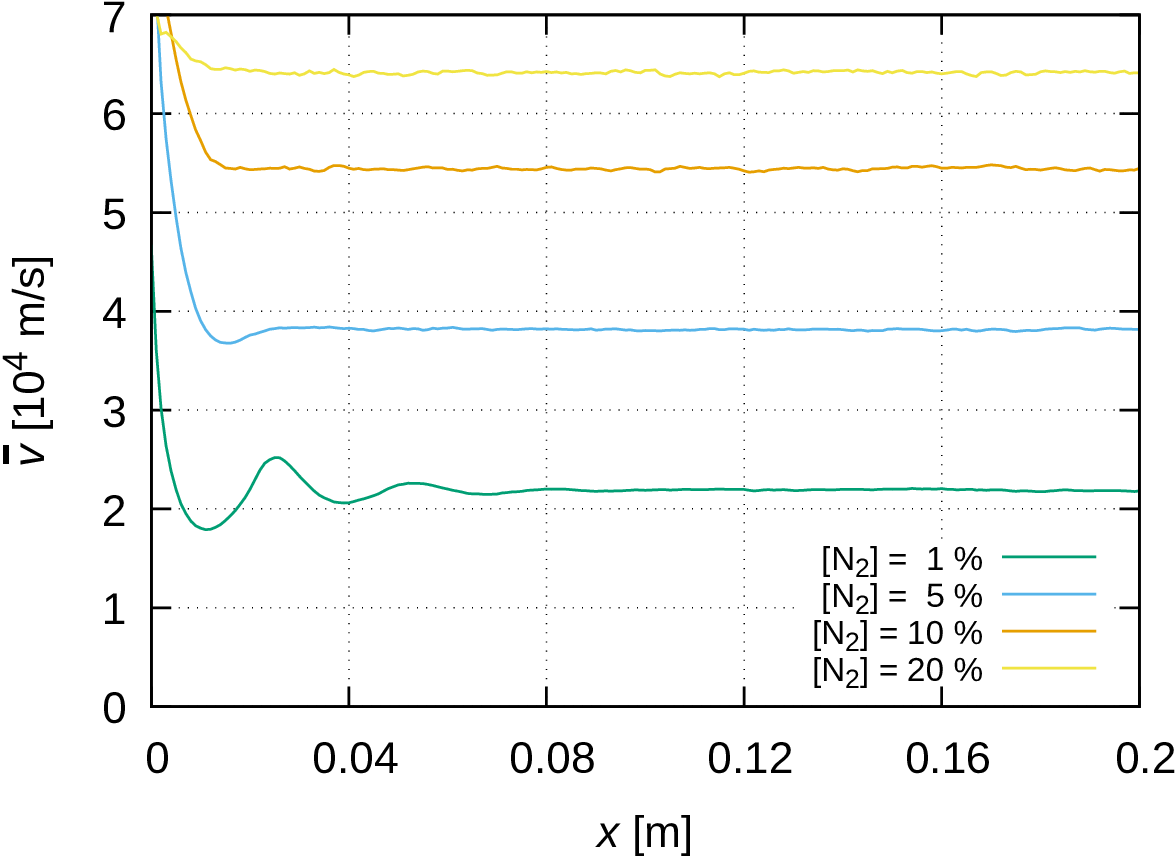}
\footnotesize(d) \includegraphics[width =0.45\textwidth]{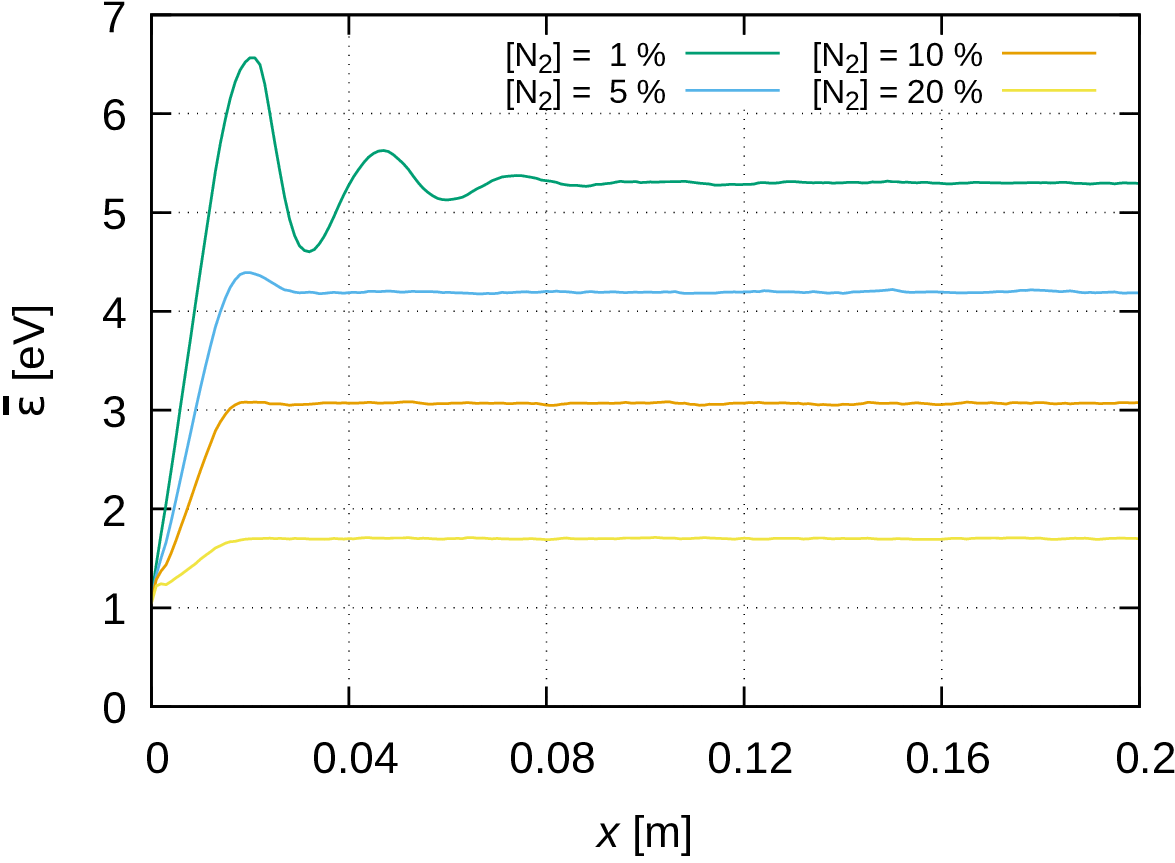}\\
\caption{Relaxation of the mean velocity (left column) and the mean energy (right column) of electrons in a homogeneous electric field. The electrons are emitted at $x$ = 0 with an initial velocity pointing into the $x$ direction and corresponding to 1 eV initial energy. $p$ = 100 Pa and $T_{\rm g} = 300$ K. (a,b) Pure Ar at various values of $E/N$, (c,d) Ar-N$_2$ mixtures at $E/N$ = 20 Td and different N$_2$ concentrations. Note that the domain shown in (c,d) is much shorter. The dashed red lines in (a,b) correspond to 20 Td, with the $x$ scale 10x zoomed.}
\label{fig:homogen}
\end{center}
\end{figure}

To illustrate the equilibration of electron swarms in a homogeneous electric field, in figure \ref{fig:homogen} we depict the mean velocity and the mean energy of the electrons for a steady state scenario when electrons are continuously emitted from an electrode at $x$ = 0 and drift in the gas. In this simulation, the electrons are emitted with an initial energy of 1\,eV and an initial velocity directed towards the $x$ direction. Such an initial velocity distribution is very clearly far from the equilibrium distribution that is expected to be nearly isotropic with a small drift component. Figures \ref{fig:homogen}(a) and (b) show the case of pure Ar. Here, the ensemble of the electrons requires rather significant "flight" lengths to acquire a steady-state mean velocity, for the whole range of $E/N$ covered. The periodic structures seen during this equilibration phase originate from repetitive energy gain - energy loss cycles of the electrons: gain occurs due to acceleration in the electric field, loss occurs primarily due to inelastic collisions. This is especially well seen in the graph of $\overline{\varepsilon}(x)$ for the 20\,Td case, for which the $x$ scale is 10x zoomed (red dashed line): here saw-tooth like patterns appear, expressing the slow energy gain and rapid energy loss. While the peaks of this function appear nearly at the same position as those of $\overline{v}(x)$, the functional forms appear to be significantly different. The oscillations of the mean velocity persist for the longest spatial domain for the 20 Td case, both at lower and higher fields we observe equilibration on a shorter length scale. 

Figures \ref{fig:homogen}(c,d) illustrates the behaviour of the swarm in Ar-N$_2$ mixtures, as a function of the N$_2$ concentration, at fixed $E/N$ = 20 Td. The equilibration of the transport takes place on a much shorter length scale, as compared to that in pure Ar. Already 1\% of N$_2$ shrinks the equilibration domain by a factor of $\sim\,10$. At higher admixture concentrations equilibrium becomes close to monotonic. This behaviour originates from the wide range of energies (due to processes with low threshold energies, e.g. vibrational channels) of inelastic loss channels in N$_2$ as compared to Ar. We can note that an increasing N$_2$ concentration results in a remarkable increase of the steady-state mean velocity and a remarkable decrease of the steady-state mean energy. 

Whenever pronounced structures in transport coefficients are seen, it is expected that in a spatially modulated electric field resonances may appear at certain conditions, as it has been recognised in several earlier studies, e.g. \cite{Nicoletopoulos}. This is indeed the foreseen behaviour in pure Ar, while N$_2$ is expected to have a converse influence on this effect.

The data presented above help setting the proper range of $E/N$ for the studies of the transport in spatially modulated fields and the timing of the data collection in the simulations (see below).

\subsection{Transport in periodically modulated electric field in argon}

\label{sec:results-ar}

\begin{figure}[ht ]
\footnotesize(a) \includegraphics[width =0.4\textwidth]{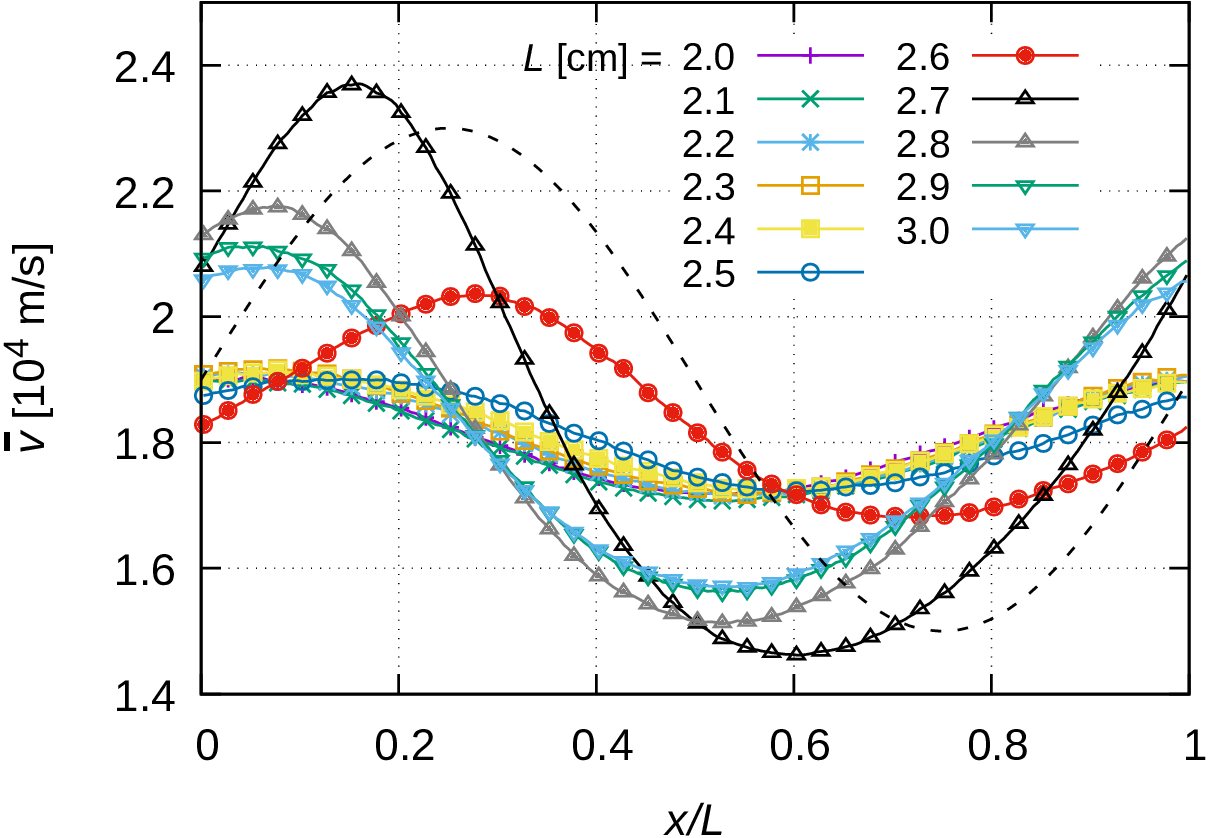}
\footnotesize(b) \includegraphics[width =0.4\textwidth]{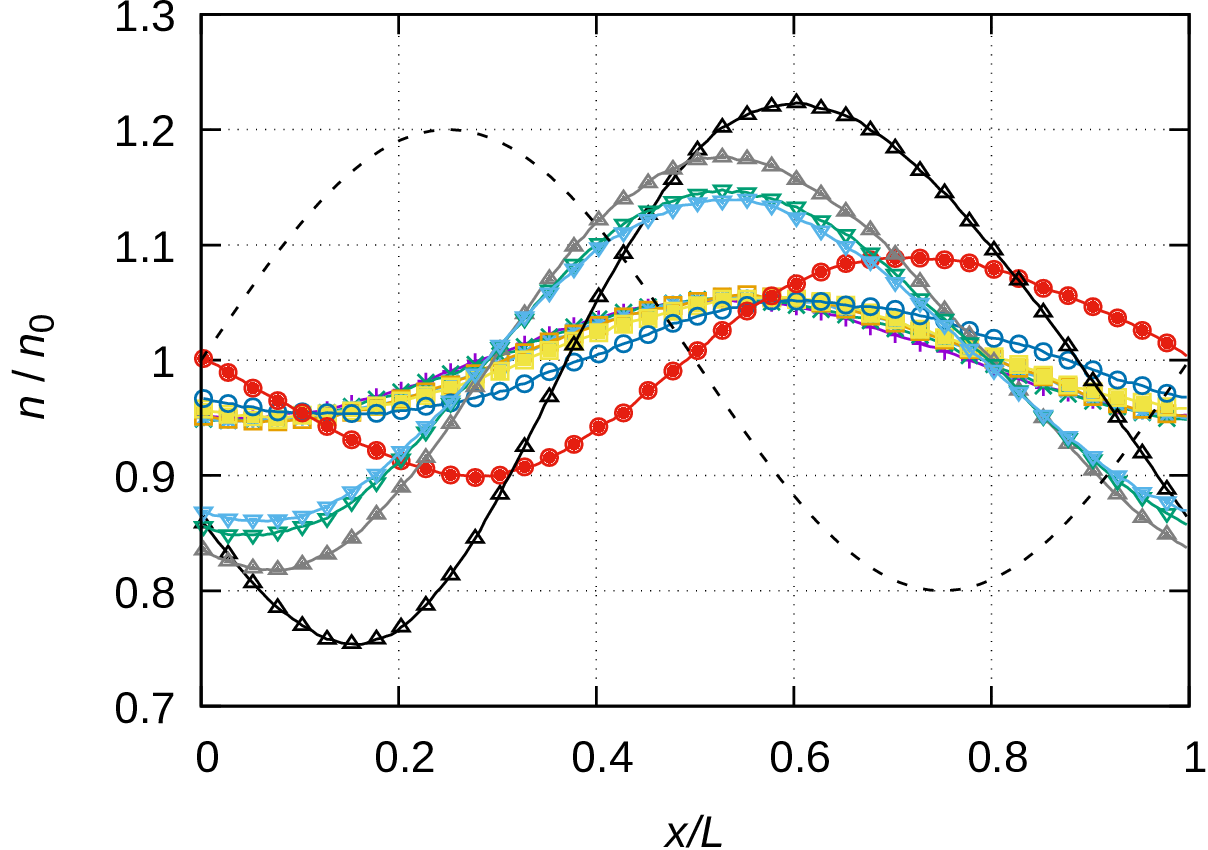}\\
\footnotesize(c) \includegraphics[width =0.4\textwidth]{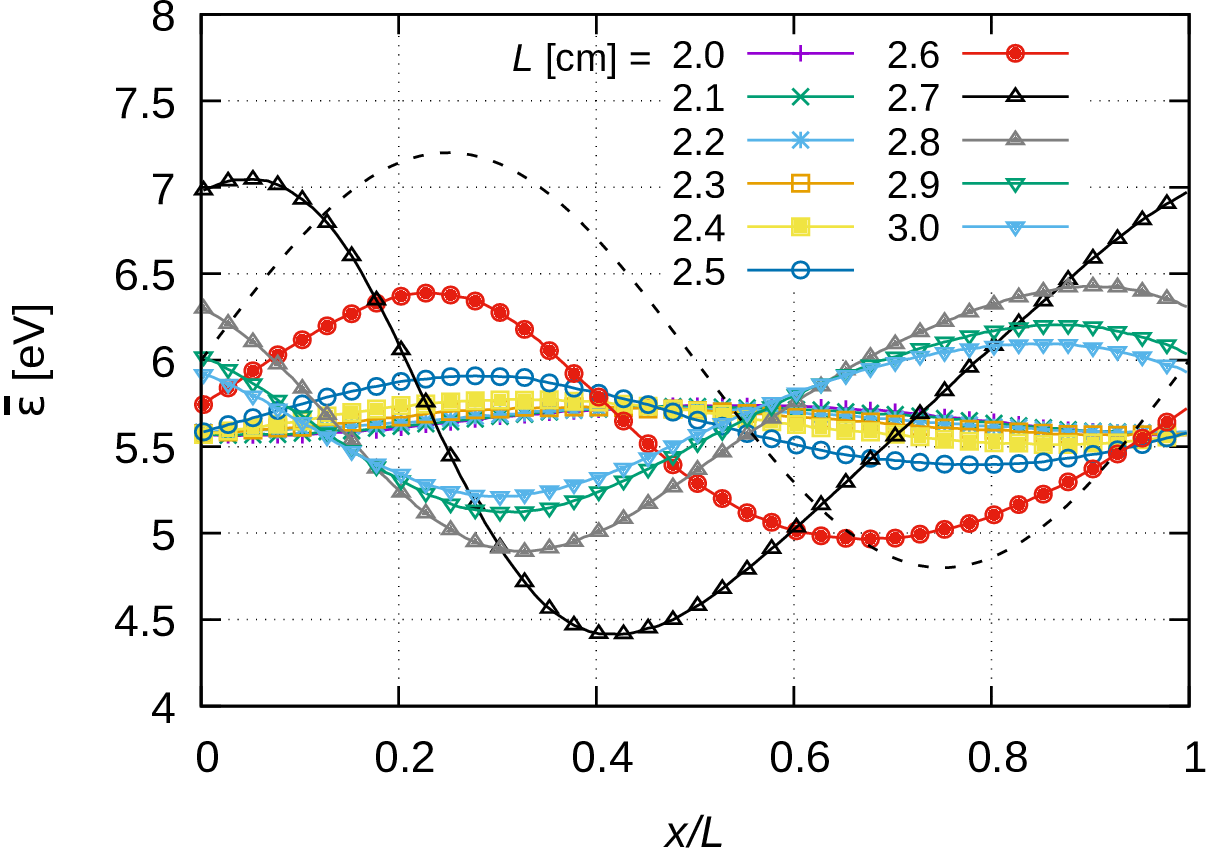}
\footnotesize(d) \includegraphics[width =0.46\textwidth]{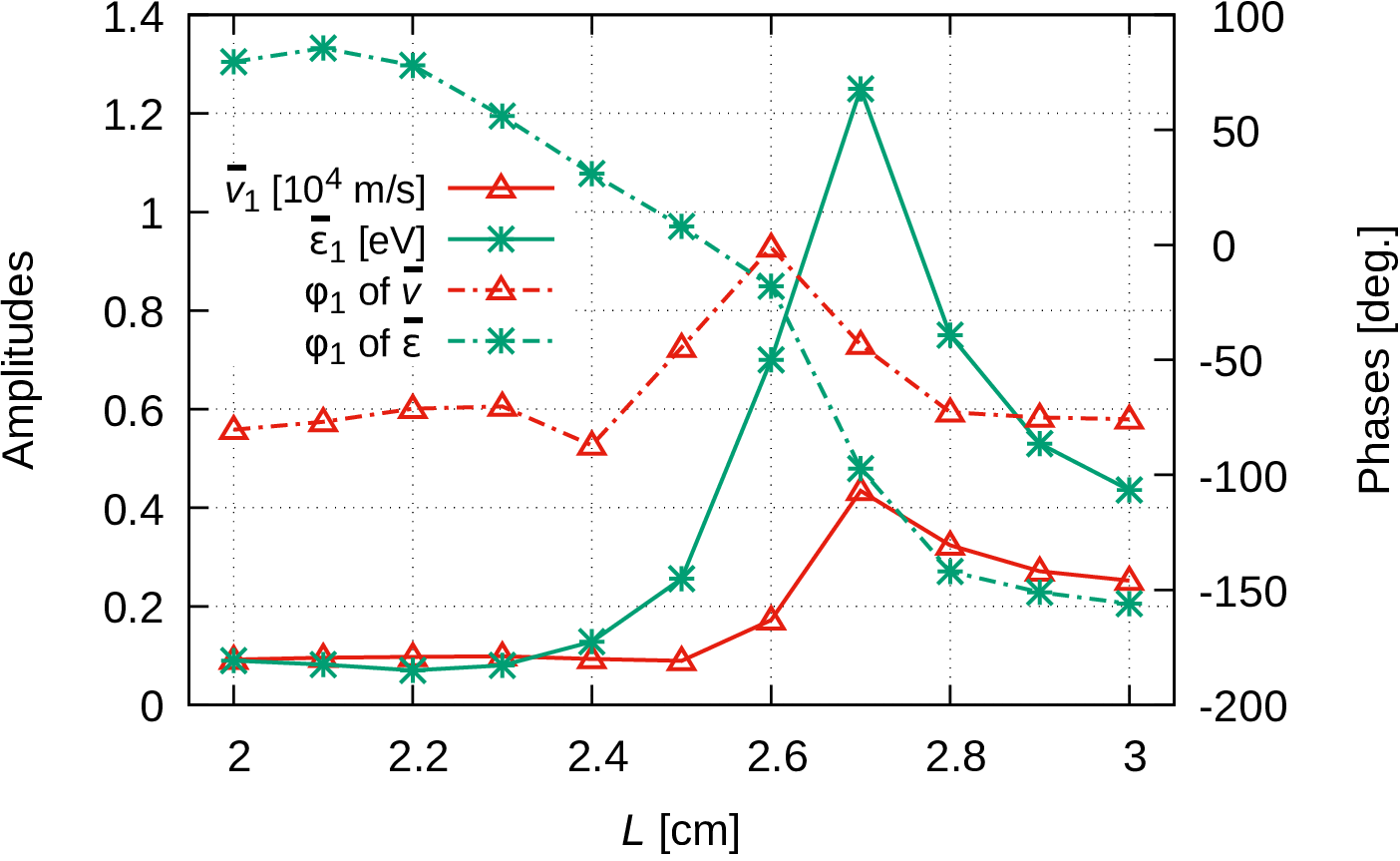}
\caption{Spatially resolved mean velocity (a), normalized density (b) and mean energy (c) of the electrons at $(E/N)_0$ = 20 Td and $M$ = 0.2, as a function of $L$. ($n_0$ is the spatial average of the electron density.) The identification of the curves in (c) is the same as in (b). The dashed black lines in each panel show the spatial variation of $E/N$, these curves are given without units. (d) Amplitudes (left scale, solid lines) and phases (right scale, chain lines) of the first ac component of the profiles of the mean velocity and the mean energy, as a function of $L$. The amplitudes of $\overline{v}_1$ and $\overline{\varepsilon}_1$ are given in units of 10$^4$ m/s and eV, respectively.}
\label{fig:mod1}
\end{figure}

Below, we present the results  for the transport properties in spatially modulated electric field for pure Ar gas. The simulations are initialized by placing $N$ = 2000--5000 electrons at random positions within the simulation box (of length $L$) with a velocity ${\bf v}_0$=0.  First, the electrons are traced for $\Delta T_1 = 900 \,\mu$s, and subsequently for an additional $\Delta T_2 = 100 \,\mu$s, during which the transport data are collected (unless stated otherwise). This timing ensures the decay of the initial transients: at 20 Td, e.g., the relaxation length seen in figure \ref{fig:homogen} is $\sim$\,1\,m and the mean stationary velocity is about $v_0 \approx 1.8 \times 10^4$ m/s, resulting in a characteristic relaxation time of $\approx$ 55 $\mu$s $\ll \Delta T_1$.  

First, we analyze the results obtained at $(E/N)_0$ = 20 Td as the longest relaxation length was observed (in figure \ref{fig:homogen}) for this value of the reduced electric field. Figure \ref{fig:mod1} shows the effect of the length of the simulation domain, $L$, for this $(E/N)_0$ and for a modulation of $M$ = 0.2. 
Panel (a) shows the mean velocity $\overline{v}(x)$, (b) the normalised electron density $n/n_0$ (where $n_0$ is the spatial average of the electron density), and (c) the mean energy $\overline{\varepsilon}(x)$. $L$ is varied between 2 cm and 3 cm, in 0.1 cm steps. Recall that the effect of $L$ translates directly to the effect of the voltage drop over the simulation cell, $U$, via eq.\,(\ref{eq:voltage}), i.e. for the present conditions the voltage is in the 9.68\,V $\leq U \leq$ 14.52\,V range.

\begin{figure}[ht ]
\begin{center}
\includegraphics[width =0.47\textwidth]{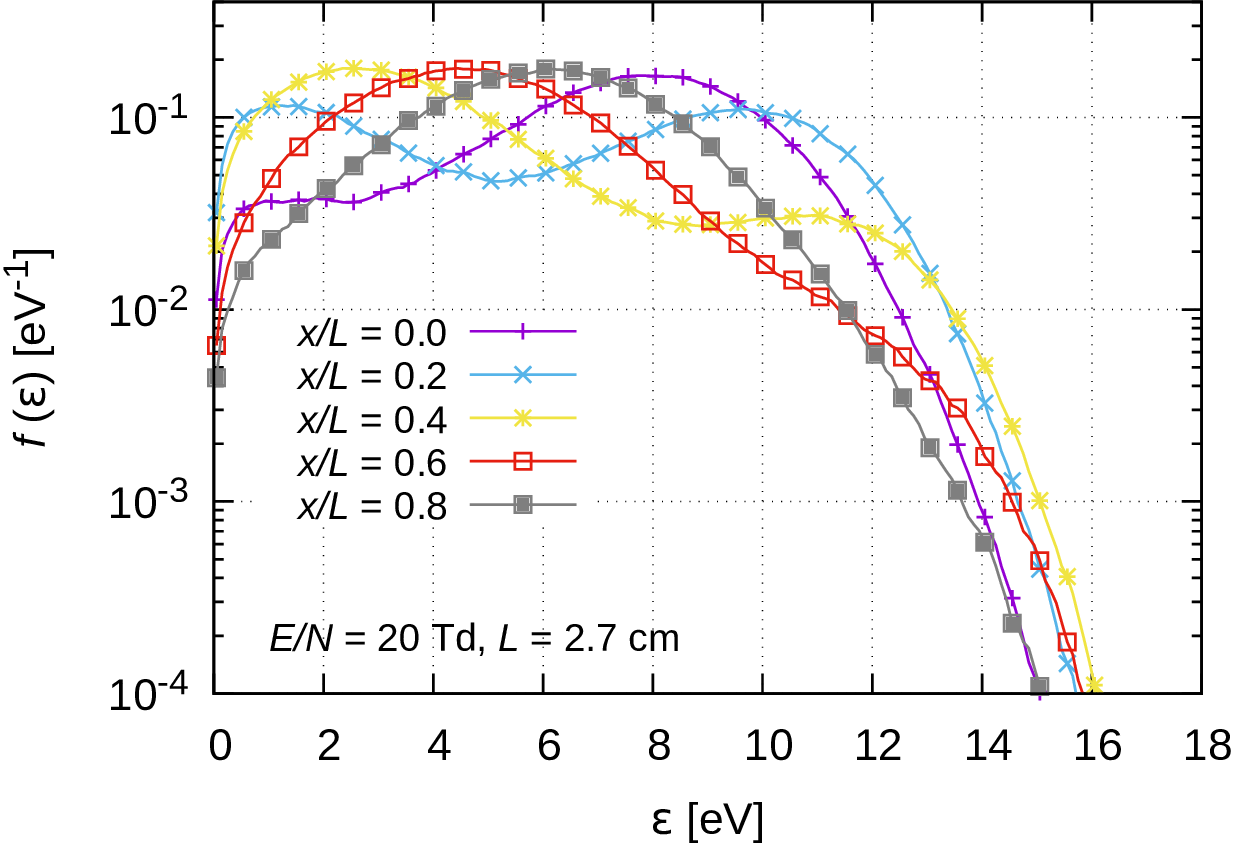}
\caption{EEDF-s at different spatial locations, at $(E/N)_0$ = 20 Td and $M$ = 0.2 and $L$ = 2.7 cm.}
\label{fig:EEDF1}
\end{center}
\end{figure}

The characteristics of the spatial profile of the quantities shown in figure \ref{fig:mod1}(a-c) (i.e. the amplitude and shape of the curves as well as the positions of their extrema) vary in a complicated manner with $L$. As regards to of $\overline{v}$, in the limit of small $L$ values we observe a weak modulation around the equilibrium value of $v_0 \approx 1.8 \times 10^4$\,m/s, with a peak close to the edge of the cell. A notable increase of the amplitude and a shift of the maximum to higher $x/L$ appear at $L$ = 2.6 cm, while the highest modulation is observed at $L$ = 2.7 cm, which is, however, accompanied by a "backward" shift of the profile. At $L >$ 2.7 cm we observe a decreasing amplitude of the profile, with maxima approaching $x/L =0$. For the given strength of modulation, $M=0.2$, the higher harmonic content of the profiles is limited. Therefore the strength of the "response" of  $\overline{v}$ to the perturbing electric field variation is characterized by the amplitude and the phase of the first Fourier component,  $\overline{v}_1$ and $\varphi_1$, in figure \ref{fig:mod1}(d). This figure confirms the visual observation of a resonance at $L$ = 2.7 cm (corresponding to $U$ = 13.07 V), where $\overline{v}_1$ exhibits a sharp peak. The phase of the profile is near $-80^\circ$ both at low and high $L$ and shows a peak at $\varphi_1 \approx 0^\circ$ at $L$ = 2.6 cm. As the system is conservative, i.e. there are no sources and losses, $n \overline{v}$ = const. holds due to flux conservation ($\nabla \cdot (n \overline{v})=0$). Therefore, the electron density obtained from the simulation (and shown in figure~\ref{fig:mod1}(b) is directly related to the mean velocity. The dependence of the spatial profile of the mean electron energy, $\overline{\varepsilon}(x)$ as a function of $L$ is similar to that of the mean velocity, as it can be seen in figure~\ref{fig:mod1}(c). The phases of the $\overline{v}(x)$ and the $\overline{\varepsilon}(x)$ profiles are, however, quite different as revealed quantitatively in figure \ref{fig:mod1}(d). The phase of the latter exhibits a monotonic decrease with the increase of $L$ and passes through $0^\circ$ at $L$ = 2.6\,cm, near the resonance.

\begin{figure}[ht ]
\begin{center}
\footnotesize(a) \includegraphics[width =0.55\textwidth]{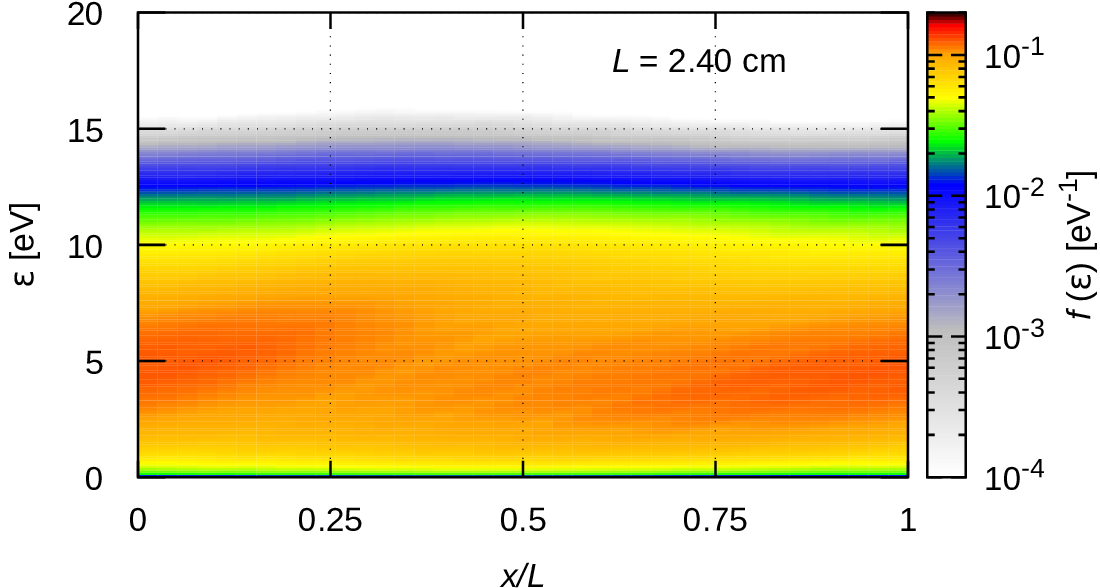}\\
\footnotesize(b) \includegraphics[width =0.55\textwidth]{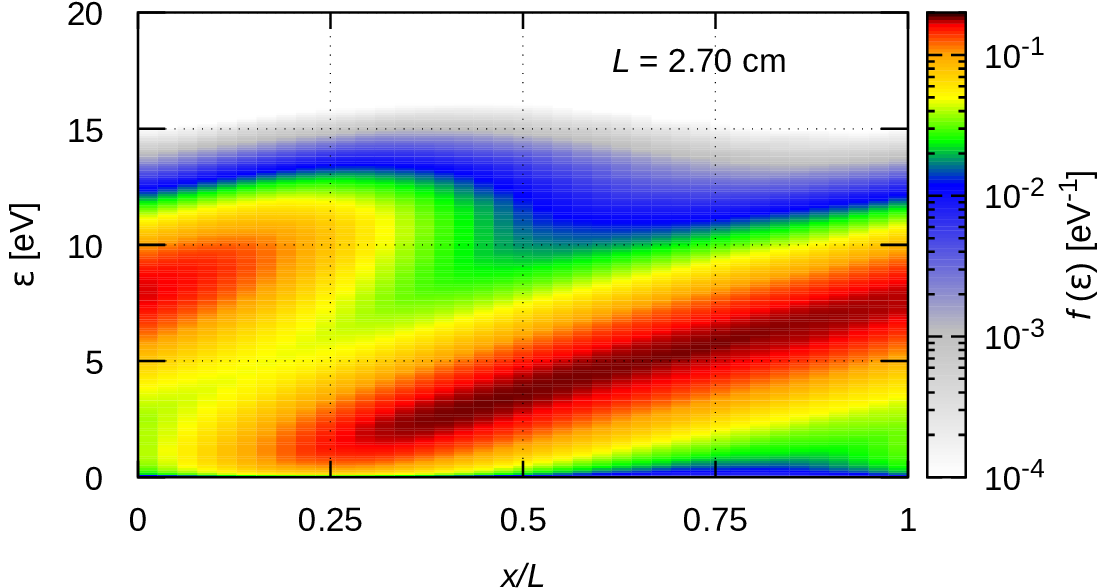}\\
\footnotesize(c) \includegraphics[width =0.55\textwidth]{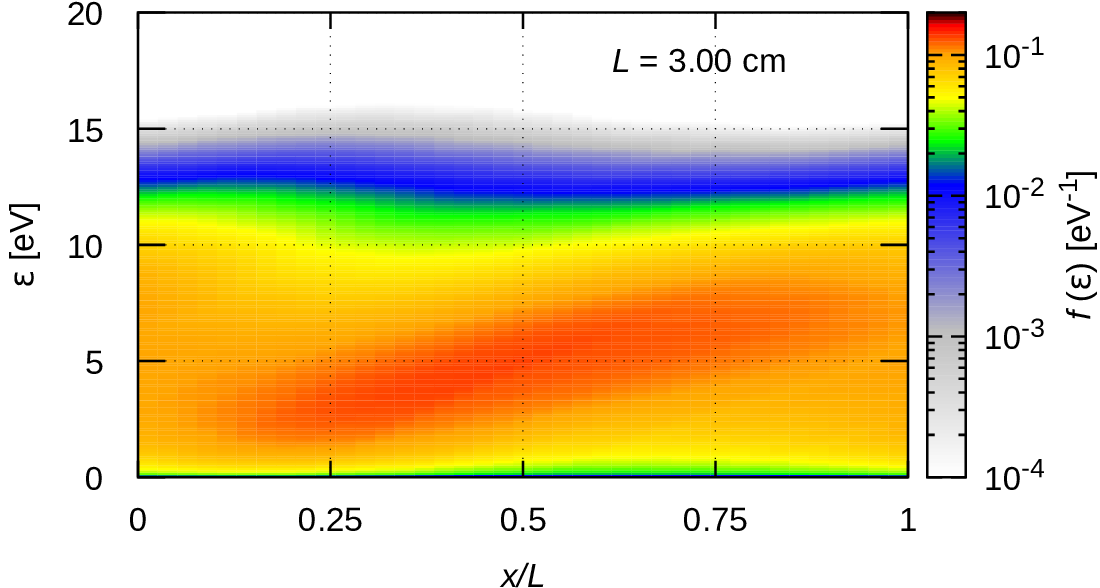}
\caption{Spatial maps of the EEDF for $(E/N)_0$ = 20 Td and $M$ = 0.2, at various values of $L$.}
\label{fig:EEDF2}
\end{center}
\end{figure}

The Electron Energy Distribution Function (EEDF) exhibits marked changes as a function of the position when $M>0$. An example of this is shown in figure \ref{fig:EEDF1} for $(E/N)_0$ = 20 Td and $M$ = 0.2, at $L$ = 2.7 cm, i.e. for the resonant case. It is remarkable that the strongest high-energy tail of the EEDF develops at the spatial position of $x/L$ = 0.4, where, actually the mean energy has a minimum (cf. figure~\ref{fig:mod1}(b)). This is not a contradiction as low-energy part of the EEDF at this position is also highly populated. On the other hand, this observation points out the importance of the whole EEDF in determining the characteristics of the transport. Revoking figure \ref{fig:LF}(b) we can note that the EEDF-s are populated at energies both below and above the energy where a sudden drop in the energy relaxation length occurs. Therefore, as pointed out in \cite{Winkler2002} the low- and high-energy parts of the electron population behave in a quite different ways at this resonance. Below the inelastic excitation threshold of Ar (11.55\,eV) $\lambda_{\rm e}$ is in the order of several cm-s while this drops to several mm-s when the energy is higher than this threshold. The long $\lambda_{\rm e}$ at low $\varepsilon$ assists the electrons to gain energy from the field, without dissipating it. The short $\lambda_{\rm e}$ at $\varepsilon > $ 11.55\,eV, on the other hand, allows the electrons to dissipate their energy quickly. Actually, at $\varepsilon =$ 11.55 eV, $\lambda_{\rm e} = $2.79\,cm. The energy accumulation for the resonant case ($L$ = 2.7\,cm) is clearly indicated by the slanted structure in panel (b) of figure \ref{fig:EEDF2} that shows the complete spatial evolution of the EEDF. Remains of this structure are also seen in figures \ref{fig:EEDF2}(a) and (c), however, these are far less pronounced.

\begin{figure}[ht ]
\begin{center}
\footnotesize(a) \includegraphics[width =0.405\textwidth]{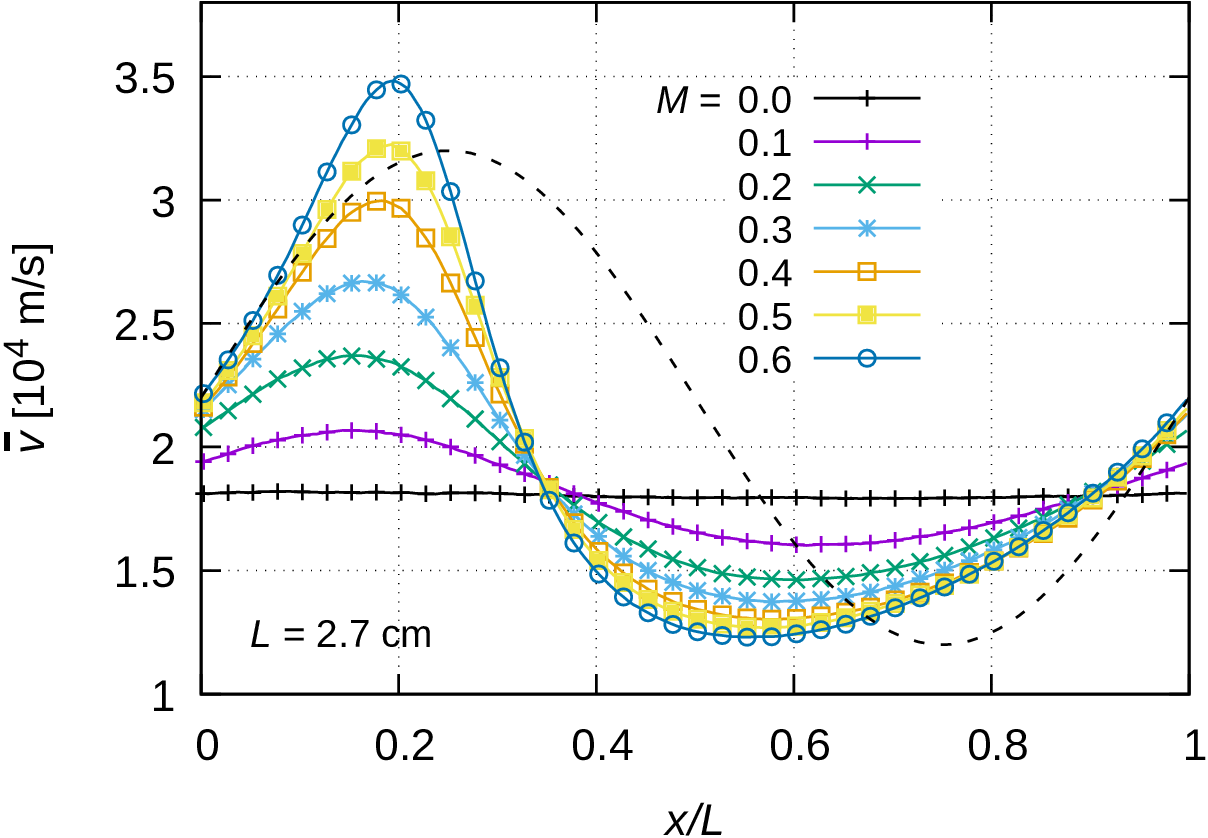}
\footnotesize(b) \includegraphics[width =0.475\textwidth]{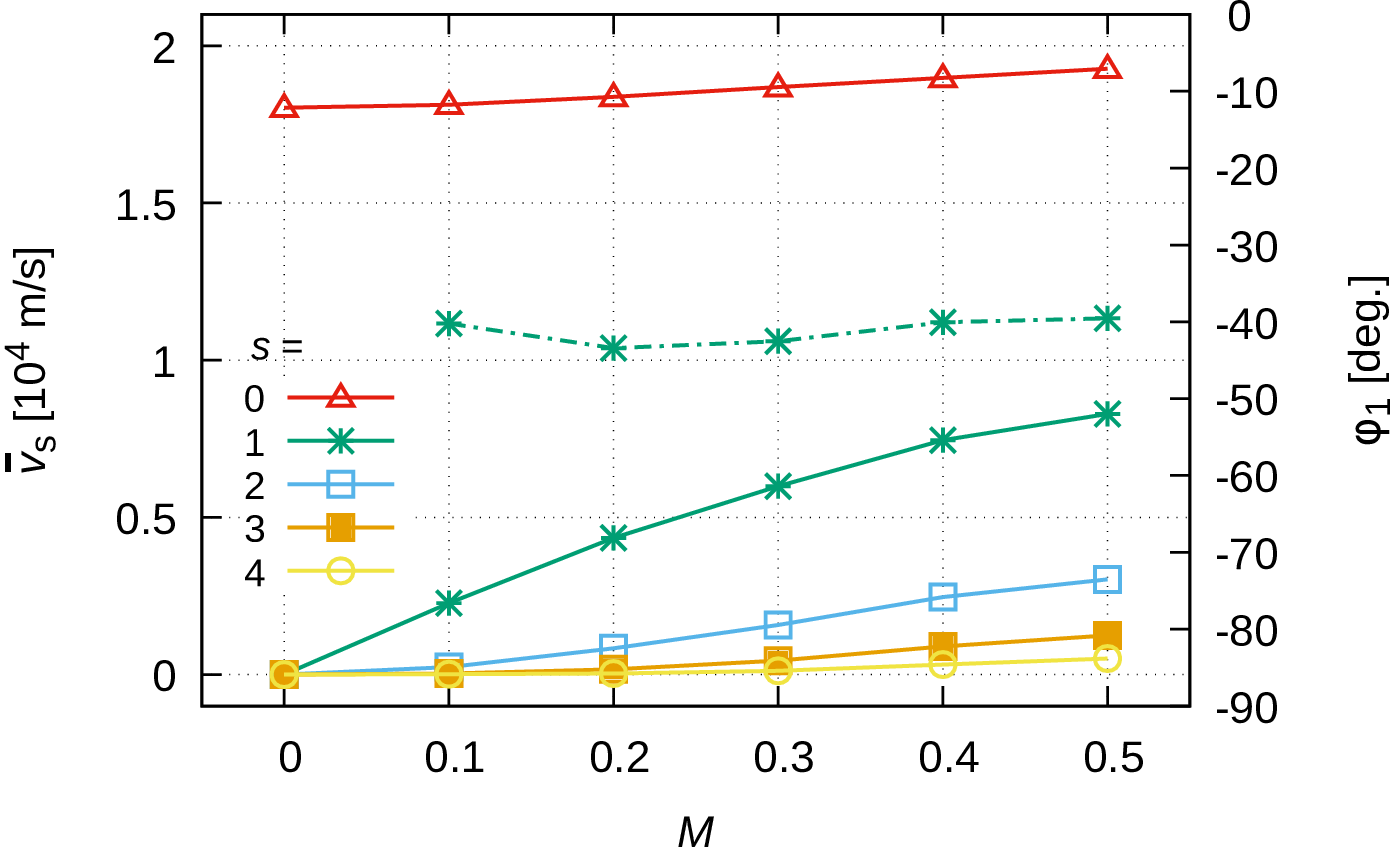}
\caption{(a) The effect of modulation depth on the spatially resolved mean velocity. The dashed black line shows the spatial variation of $E/N$, this curve is given without units. (b) Harmonic composition of $\overline{v}(x)$ (solid lines, left scale) and the phase of the first ac component, $\varphi_1$ (chain curve, right scale) as a function of the modulation depth of the electric field. $L$ = 2.7 cm and $(E/N)_0$ = 20 Td.}
\label{fig:m_effect}
\end{center}
\end{figure}

The effect of the modulation depth, $M$, on the spatial variation of the mean velocity is depicted in figure \ref{fig:m_effect}(a). At the lower values of $L$, the $\overline{v}(x)$ curves are nearly harmonic, an increasing anharmonicity can be observed with increasing modulation. Figure \ref{fig:m_effect}(b) shows the harmonic composition of $\overline{v}(x)$ as a function of $M$ for the case of $L=$ 2.7 cm and $(E/N)_0$ = 20 Td. Besides the harmonic amplitudes, $\overline{v}_s$, the phase of the principal component of the "response", $\varphi_1$ is also shown. At low modulation, only $\overline{v}_1$ differs significantly from zero, however, with increasing $M$ the harmonic content increases. The phase is $\varphi_1 \approx -40^\circ$ for all $M$. We can note that the $s=0$ component slightly increases with $M$, i.e. the "dc component" $\overline{v}_{s=0}$ of $\overline{v}(x)$, which is the spatially averaged velocity of the electrons, increases as an effect of the modulation. 

\begin{figure}[ht ]
\begin{center}
\footnotesize(a) \includegraphics[width =0.45\textwidth]{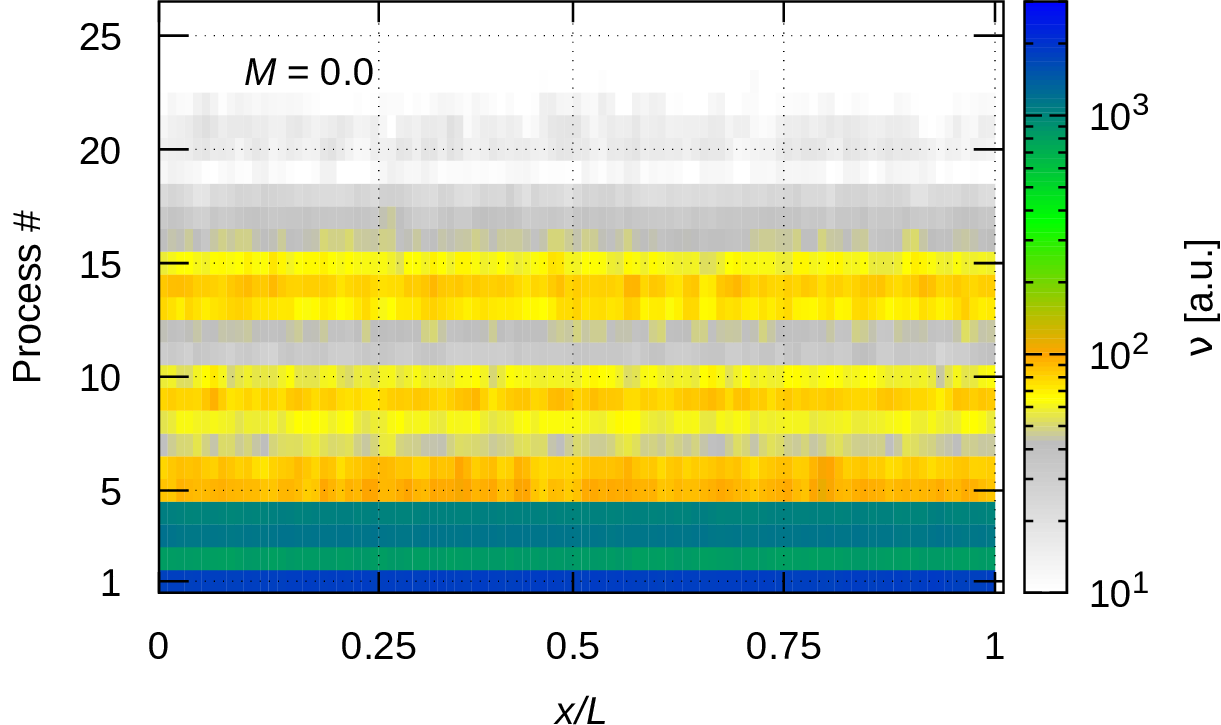}
\footnotesize(b) \includegraphics[width =0.45\textwidth]{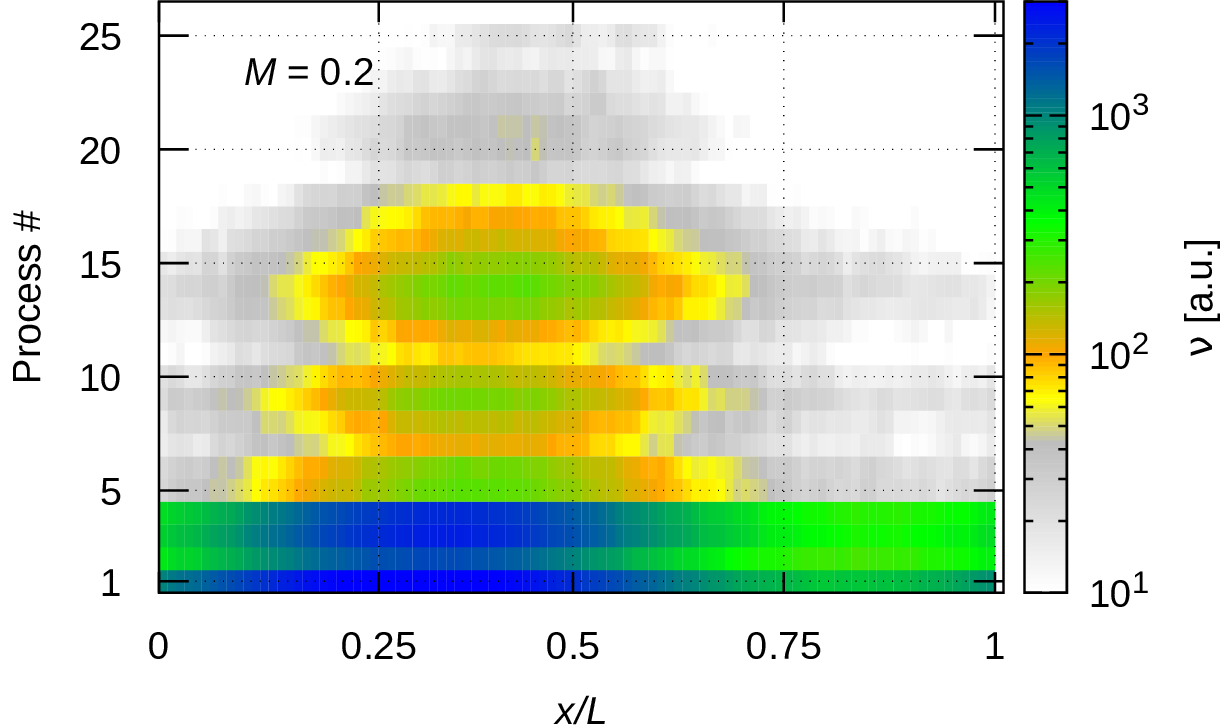}
\caption{The spatial distribution of collision frequency (in arbitrary units) of individual inelastic collision processes at $(E/N)_0$ = 20 Td and $L$ = 2.7 cm ($U$ = 13.07 V), for (a) $M$ = 0.0 and (b) $M$ = 0.2. (Note, that for these conditions ionisation (process \#26) is negligible.)}
\label{fig:mod_matrix}
\end{center}
\end{figure}

Counting the number of the different electron-Ar atom reactions spatially resolved allows construction of a matrix that shows the collision frequencies associated with the various collision processes. The computational results are shown in figure \ref{fig:mod_matrix} for the case of $(E/N)_0$ = 20 Td and $L$ = 2.7 cm, obtained at $M$ = 0.0 (panel a) and  $M$ = 0.2 (panel b). In these plots, excitation processes are identified by numbers 1...25, ionisation is process 26. The number of elastic collisions is orders of magnitude higher, thus this (process 0) is omitted from the plots. The highest excitation rates are observed (in both cases) for the four lowest excited levels of Ar, for processes 1--4, corresponding to excitation to the 1s$_5$, 1s$_4$, 1s$_3$, and 1s$_2$ levels (Paschen notation), respectively. The energy of these levels is between 11.55 eV and 11.83 eV. At $M=0$, significant rates are also observed for the whole domain for processes 5 and 6 (2p$_{10}$ and 2p$_{9}$ levels), 8--10 (2p$_{7,6,5,4}$ levels), and 13--15 (2p$_{1}$, 3d$_{5,6}$ and 3d$_{3}$ levels). The excitation threshold of the latter is 13.90 eV. Ionisation (process 26) is not present with an appreciable rate, justifying our approach of treating this process as a conservative one, for the rare events of occurrence. 

\begin{figure}[ht ]
\footnotesize(a)\includegraphics[width =0.43\textwidth]{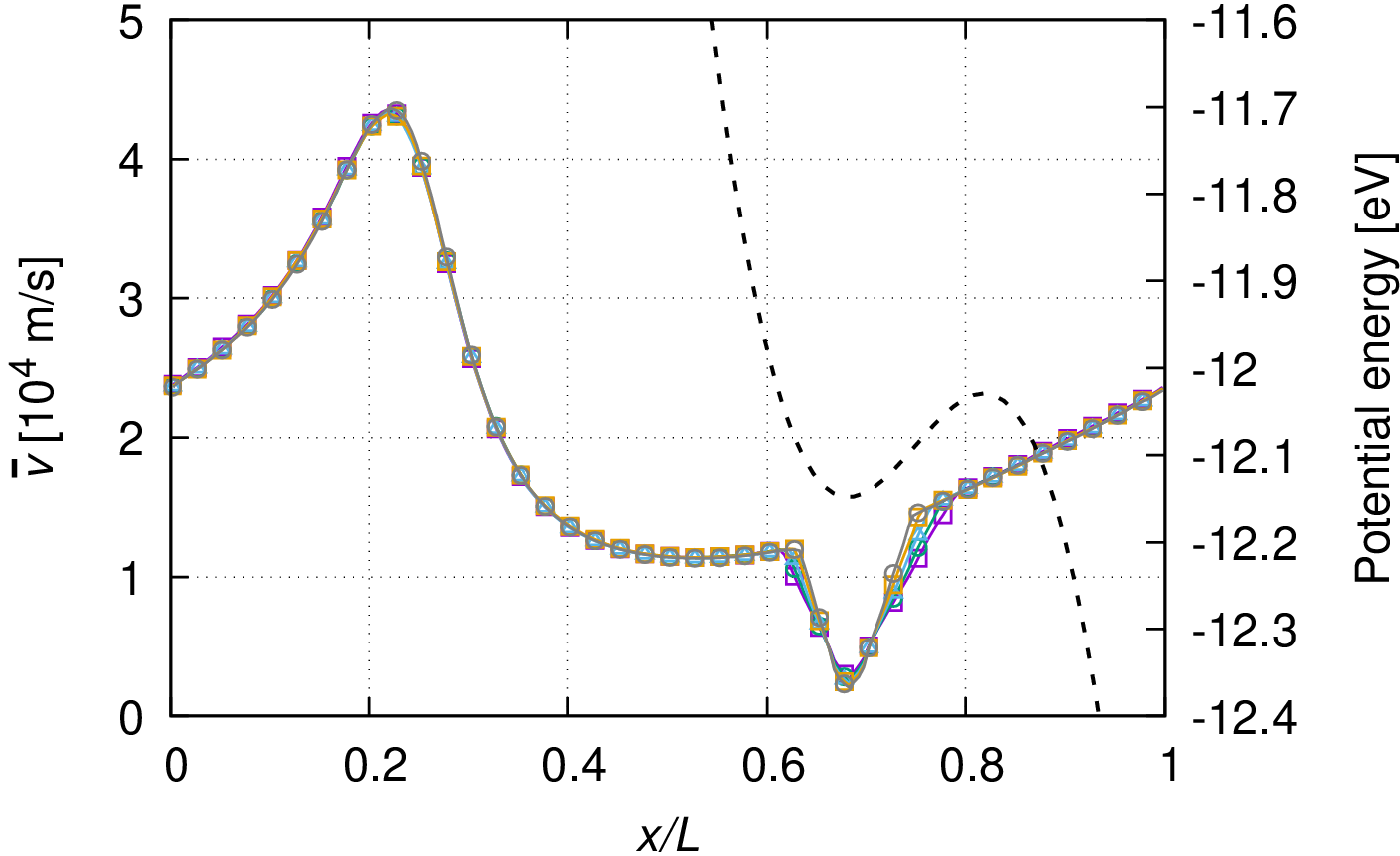}
\footnotesize(b)\includegraphics[width =0.43\textwidth]{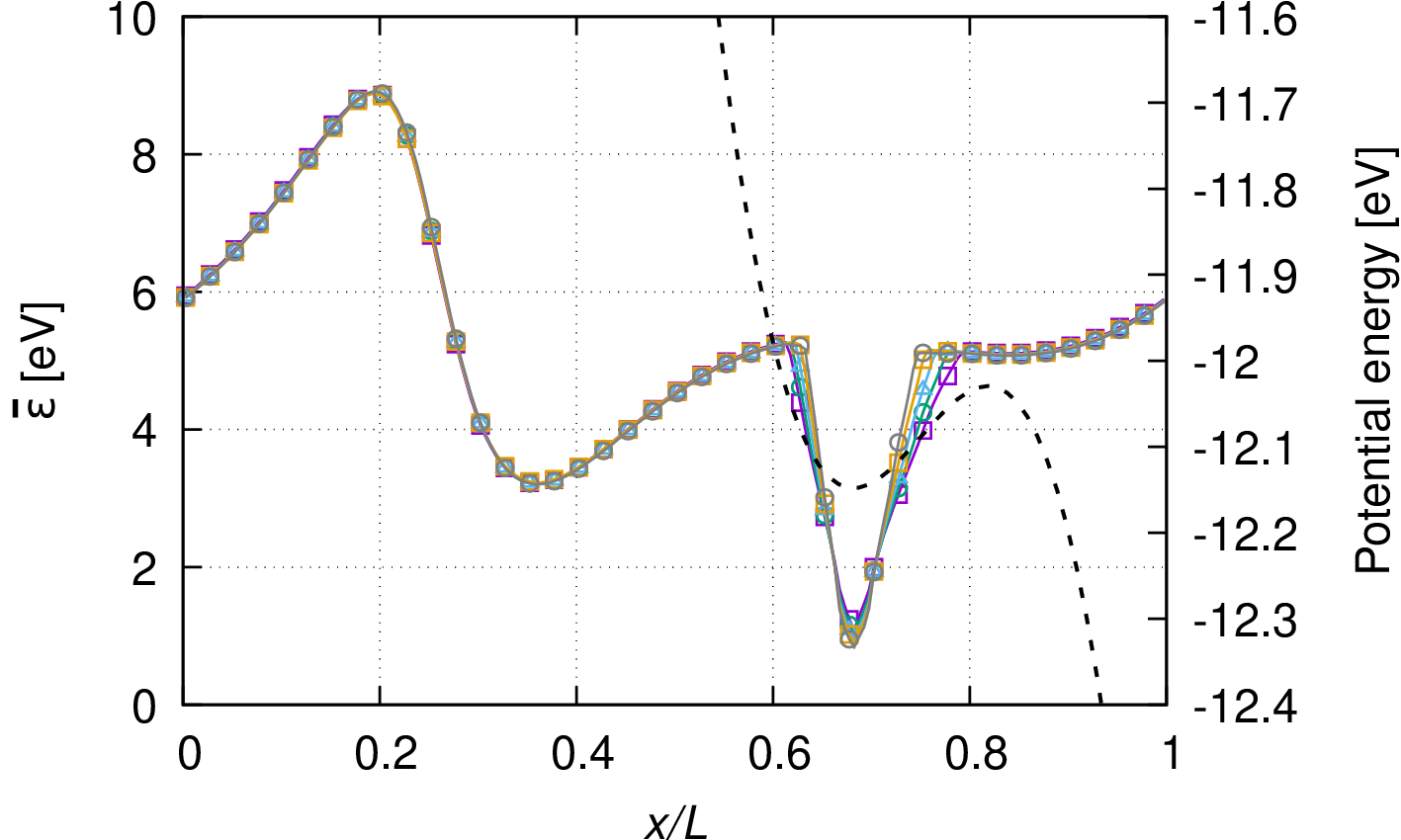}\\
\footnotesize(c)\includegraphics[width =0.43\textwidth]{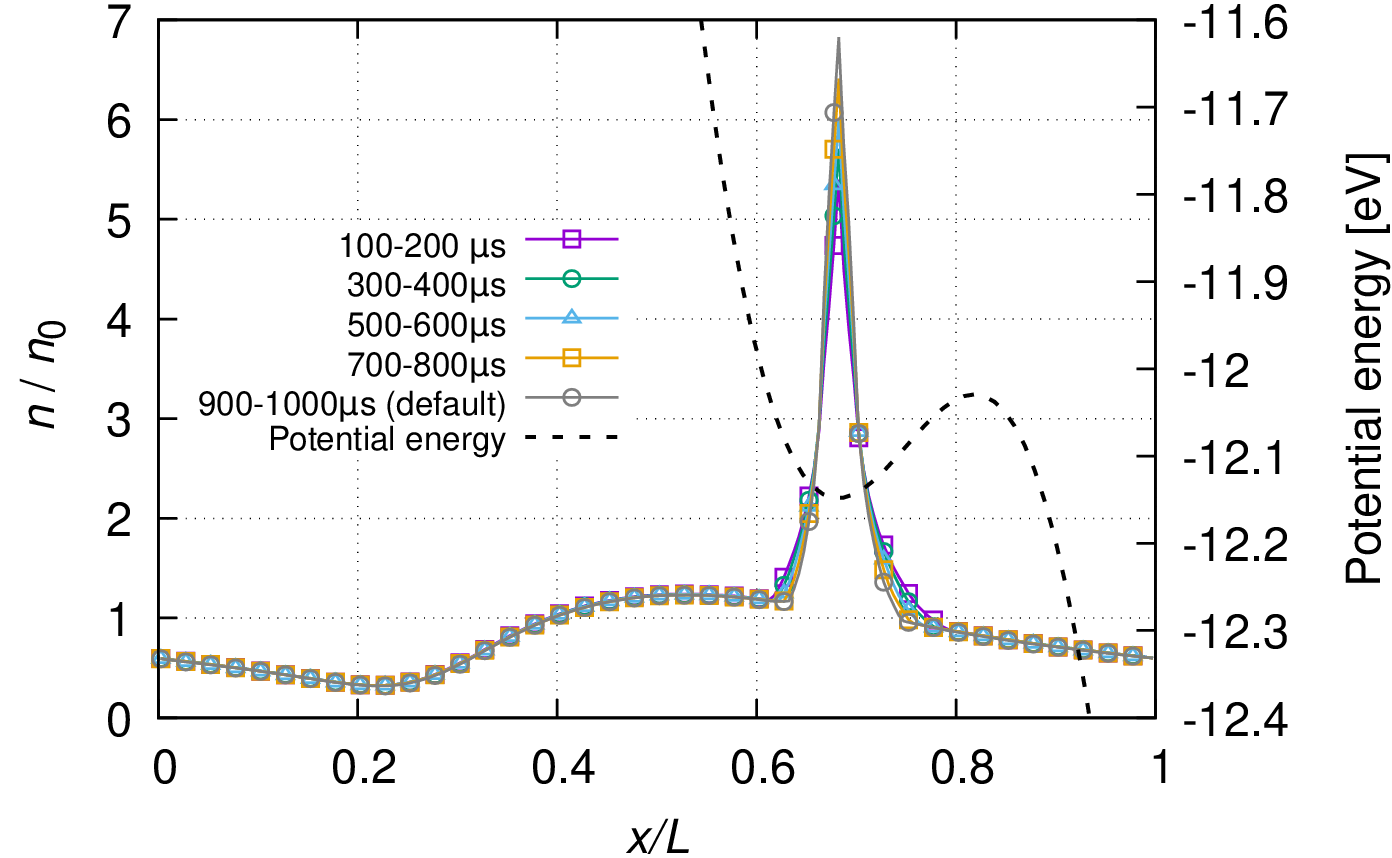}
\footnotesize(d)\includegraphics[width =0.38\textwidth]{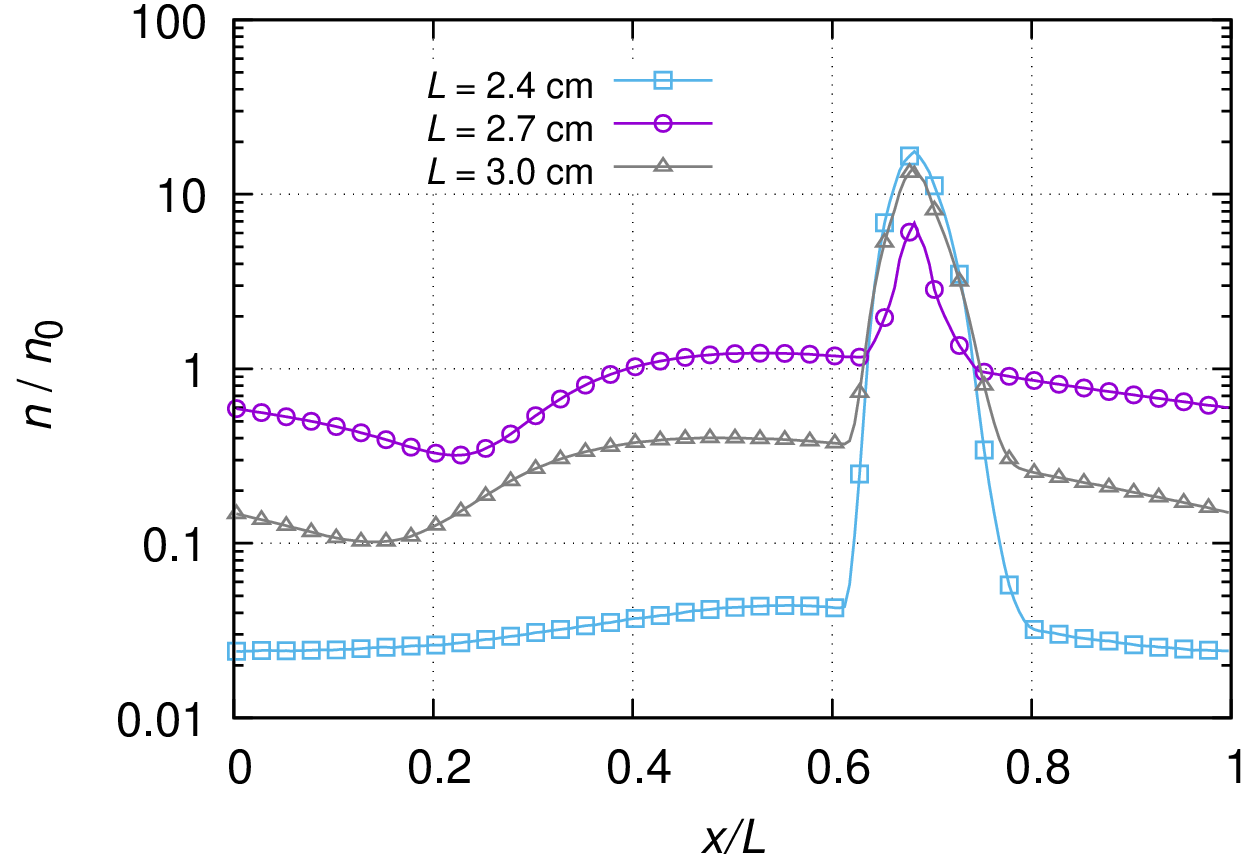}
\caption{Spatially resolved mean velocity (a), mean energy (b) and normalised density (c) of the electrons at $(E/N)_0$ = 20 Td and $L$ = 2.7 cm, for $M$ = 1.1. The curves correspond to different data collection time windows, as indicated. The legend shown in (c) is valid for panels (a) and (b), too. Note the slow evolution of the features with time within the domain of the reversed field, centered around $x/L$ = 0.75. The dashed black lines show the potential energy of the electrons, with zero value set at $x$ = 0. (d) The dependence of the normalised electron density on $L$ at $(E/N)_0$ = 20 Td and $M$ = 1.1. These data were collected in the default, 900-1000 $\mu$s time window.
}
\label{fig:extreme}
\end{figure}

At $M=0.0$ (figure \ref{fig:mod_matrix}(a)) the spatial distribution of the excitation events is homogeneous, as expected. At $M=0.2$, however, the modulation of the electric field causes a major perturbation to the excitations \cite{Aleksandrov}. We find that (i) the majority of excitation events is concentrated within the $0.25 \leq x/L \leq 0.6$ spatial domain with a pronounced maximum near $x/L = 0.4$, where the most notable high energy tail for the EEDF was found for the same conditions, (cf. figure \ref{fig:EEDF1}) and (ii) the acceleration of the electrons in the modulated field also opens excitation channels with higher thresholds: in figure \ref{fig:mod_matrix}(b) we observe processes with appreciable rates up to \#18, which corresponds to levels 3d$_{1}$" and 2s$_5$, with a threshold energy of 14.06 eV \cite{Hayashi}.

While in the cases described before, the  modulation was kept at moderate levels, here we briefly examine the case of higher modulation, when the electric field changes sign within a certain domain (at $M>1$). As the presence of a region with a reversed electric field gives rise to a potential well, electrons can accumulate within these regions. For  $(E/N)_0$ = 20 Td, $L$ = 2.7 cm, and $M$ = 1.1, e.g., the depth of this potential well is $\approx 0.1$\,V. As electrons may undergo such collisions in these regions when their remaining kinetic energy is less than 0.1\,eV, such electrons will be trapped as their energy cannot increase anymore to overcome the barrier. Consequently, after a sufficiently long time all electrons are expected to be trapped in our simulation. 

While the presence of the reversed field bears some similarity with the case of the negative glow region of dc glow discharges \cite{reversed,FR1,FR2}, in that setting fast electrons arriving from the sheath can interact with the trapped population and can eventually increase the energy of some of the electrons, enabling them this way to get released from the trap. This effect is the result of Coulomb collisions, which are however, not included in the present simulation. Nonetheless, our simulations can follow the time dependence of the trapping phenomenon. This is illustrated in figure \ref{fig:extreme}(a-c), where the results (for the mean velocity, mean energy and normalised density) are shown as a function of time. The data were collected in 100$\mu$s-wide time windows at different start times, as indicated. The graphs indeed exhibit pronounced structures around $x/L$ = 0.75, where the reversed electric field peaks. They show that a slow change follows after an initial high rate of trapping. From this it follows that the spatial modulation of $E$ actually slows down the trapping process by "moving" most of the inelastic collisions from a random distribution to spatial positions that exclude the domain of reversed field, as shown in figure \ref{fig:mod_matrix}. In accordance with this it is also interesting to note that the least significant trapping for otherwise same conditions is observed for the resonant case of $L$ = 2.7\,cm, as revealed in figure \ref{fig:extreme}(d). For other values of $L$, we observe a much more significant collection of the electrons within the region of the field reversal and more significant depletion of the density outside this domain. As electrons are less likely to undergo inelastic collisions in the regions with reversed electric field, complete trapping takes place on a time scale much longer than accessible by our simulations. Inclusion of the Coulomb collisions \cite{Kushner} and/or thermal contribution of the background gas are clearly necessary to model correctly the stationary state of our system at such high modulations.

\begin{figure}[ht ]
\begin{center}
\footnotesize(a) \includegraphics[width =0.45\textwidth]{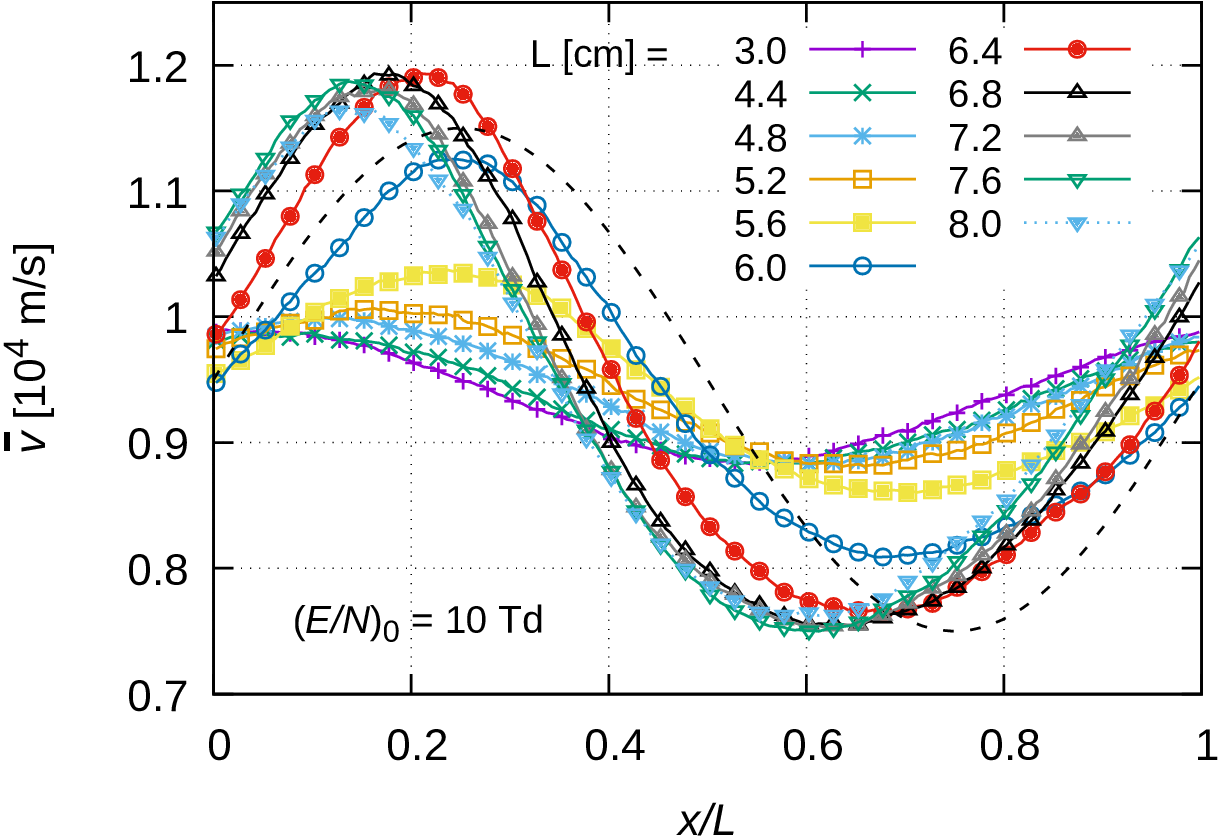}\\
\footnotesize(b) \includegraphics[width =0.45\textwidth]{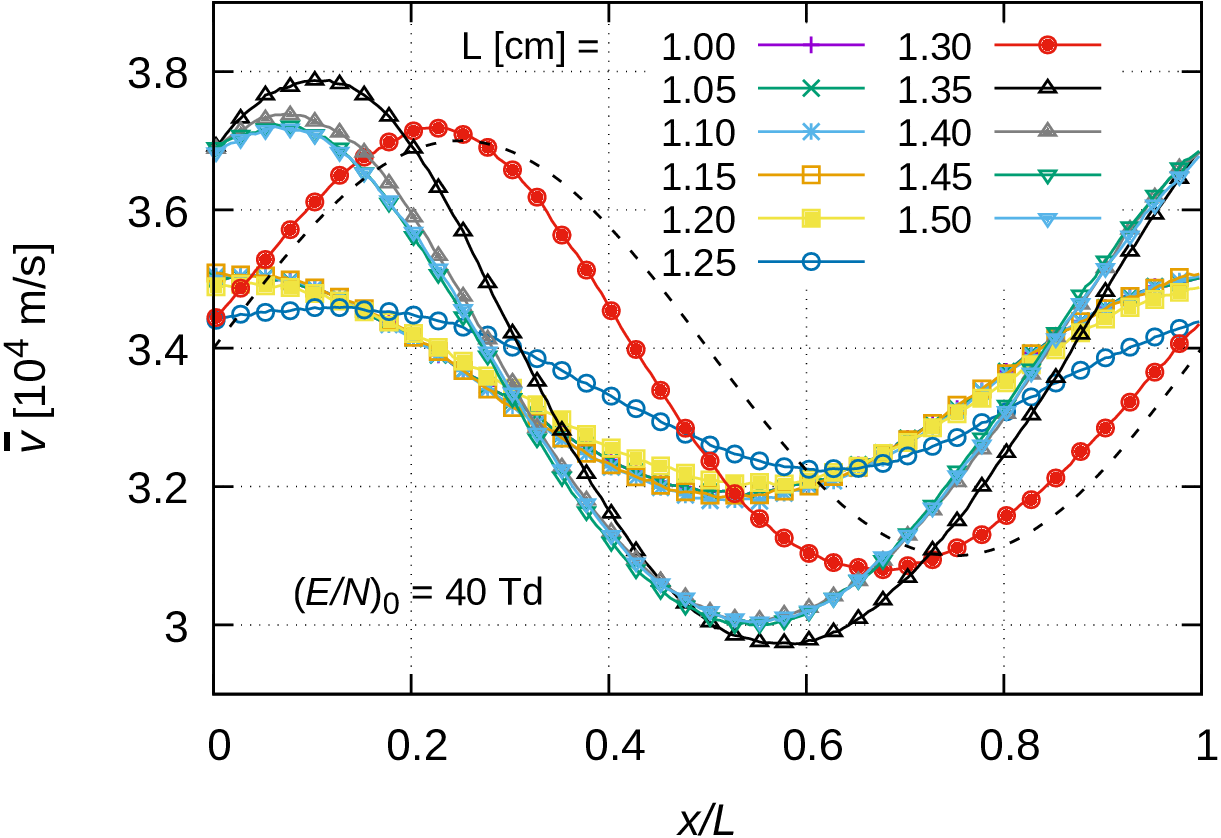}\\
\caption{Mean electron velocity at different $(E/N)_0$ values: (a) 10 Td cm, (b) 40 Td. The dashed black lines in both panels show the spatial variation of $E/N$, these curves are given without units.}
\label{fig:en}
\end{center}
\end{figure}

\begin{figure}[ht ]
\begin{center}
\includegraphics[width =0.47\textwidth]{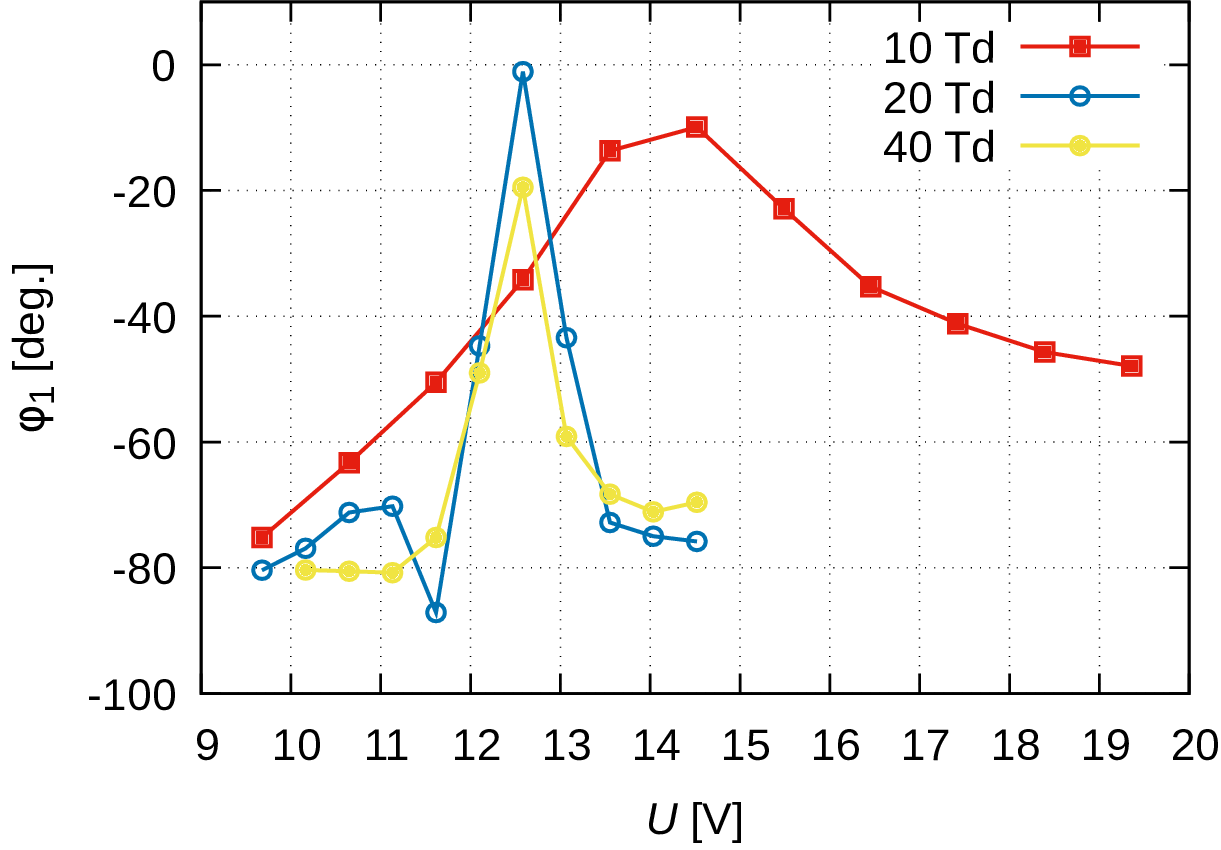}
\caption{The phase of the first ac component of the mean velocity at the $(E/N)_0$ values indicated, as a function of the voltage drop (related to the other parameters via eq.(\ref{eq:voltage})) over the computational cell. $M$ = 0.2.}
\label{fig:phase_en}
\end{center}
\end{figure}

After presenting the results obtained for $(E/N)_0$ = 20 Td, we analyse cases with lower and higher values, $(E/N)_0$ = 10 Td and 40 Td. The results for these cases are shown in figures \ref{fig:en}(a) and (b), respectively. At $(E/N)_0$ = 10 Td, the profile of the mean velocity changes smoothly with increasing $L$, without any significant resonance. The phase of the first ac component $\overline{v}_1$, changes however, in the same way as it was found previously for 20 Td. This is presented in figure \ref{fig:phase_en}, where $\varphi_1$ is shown as a function of the voltage drop $U$ (in order to make the data obtained at different $(E/N)_0$ values comparable). The maximum of the phase occurs at a higher voltage drop (higher $L$) compared to the higher fields. At $(E/N)_0$\,=\,40 Td, the behaviour of the phase is very similar to that at 20 Td (a strong peak at $\approx 12.5$\,V). The inspection of the amplitude of $\overline{v}(x)$ in figure \ref{fig:en}(b) does not show a strong resonance unlike in the case of 20 Td. This may be explained by the higher number of inelastic loss channels at an expanded energy range of the electrons at the higher accelerating field. Note, that at $E/N$\,=\,40 Td a faster spatial relaxation was found also in the homogeneous field, as compared to 20 Td, see section \ref{sec:reults-homog}.

\subsection{Transport in periodically modulated electric field in Ar-N$_2$ mixtures and in N$_2$}

\label{sec:mix}

As discussed above, the pronounced response of the electron transport parameters on the spatial modulation of the accelerating electric field in argon gas is due to the fact that the number of energy loss channels is limited (excitation predominantly occurs to a few excited states as confirmed by the results presented in the previous section). As in a molecular gas, like N$_2$ the possible values of the energy loss in a collision span a much wider domain as compared to atomic gases, the response of the system to the modulated electric field is expected to diminish when even small amounts of molecular gases are added to Ar.

\begin{figure}[ht ]
\begin{center}
\footnotesize(a)\includegraphics[width =0.45\textwidth]{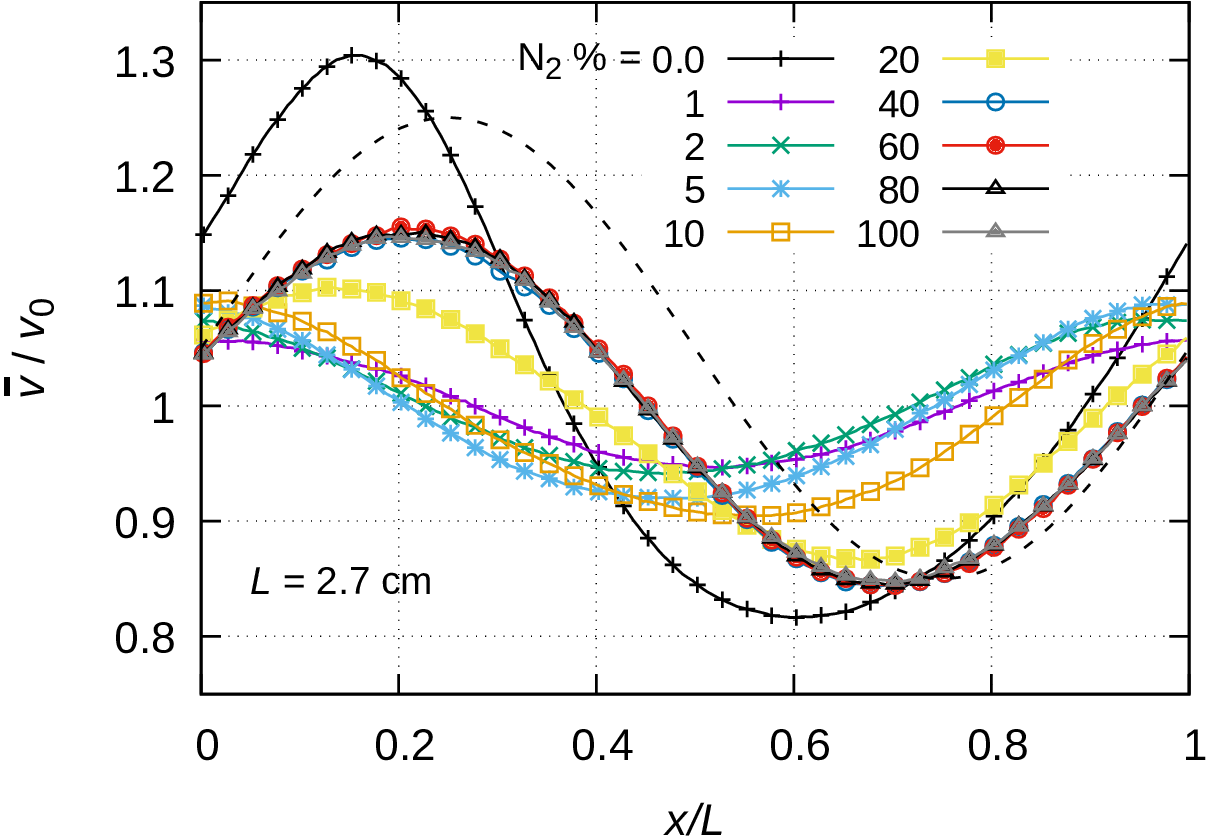}
\footnotesize(b)~~~~~\includegraphics[width =0.45\textwidth]{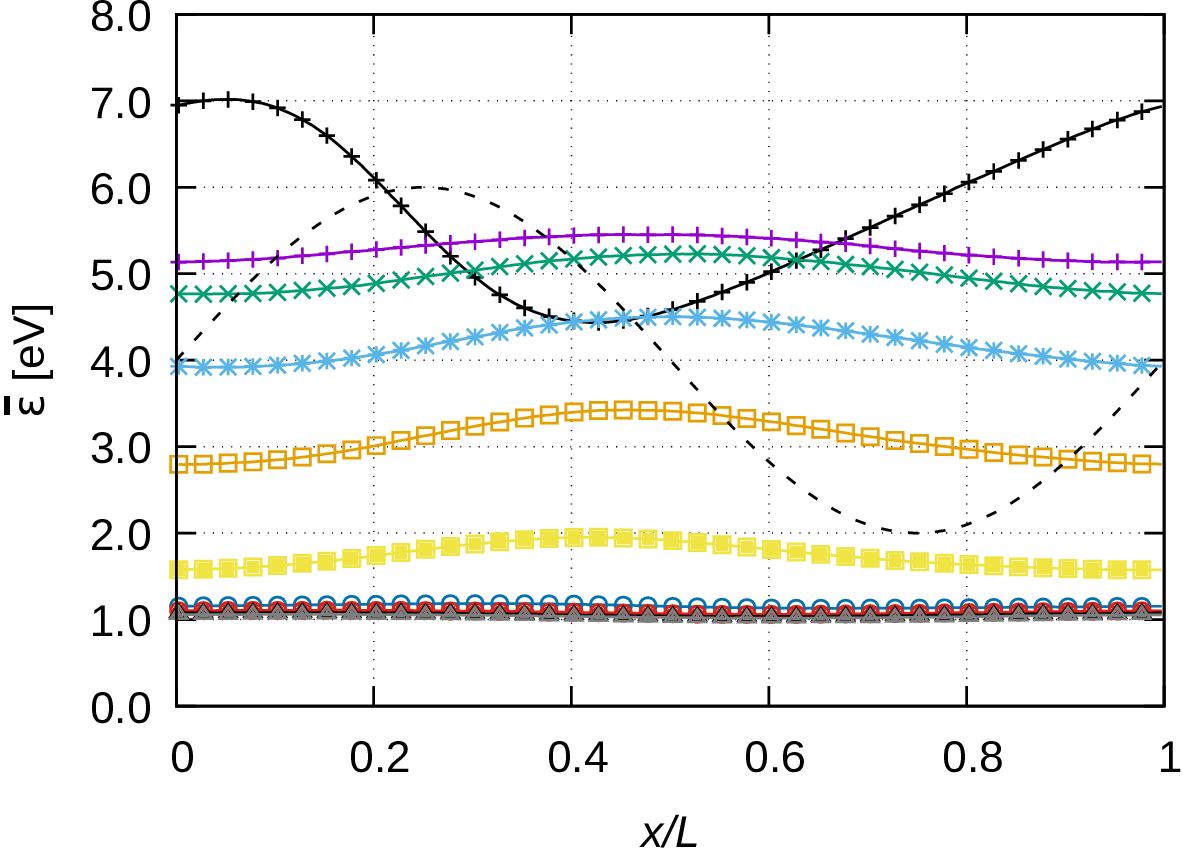}\\
\footnotesize(c) \includegraphics[width =0.45\textwidth]{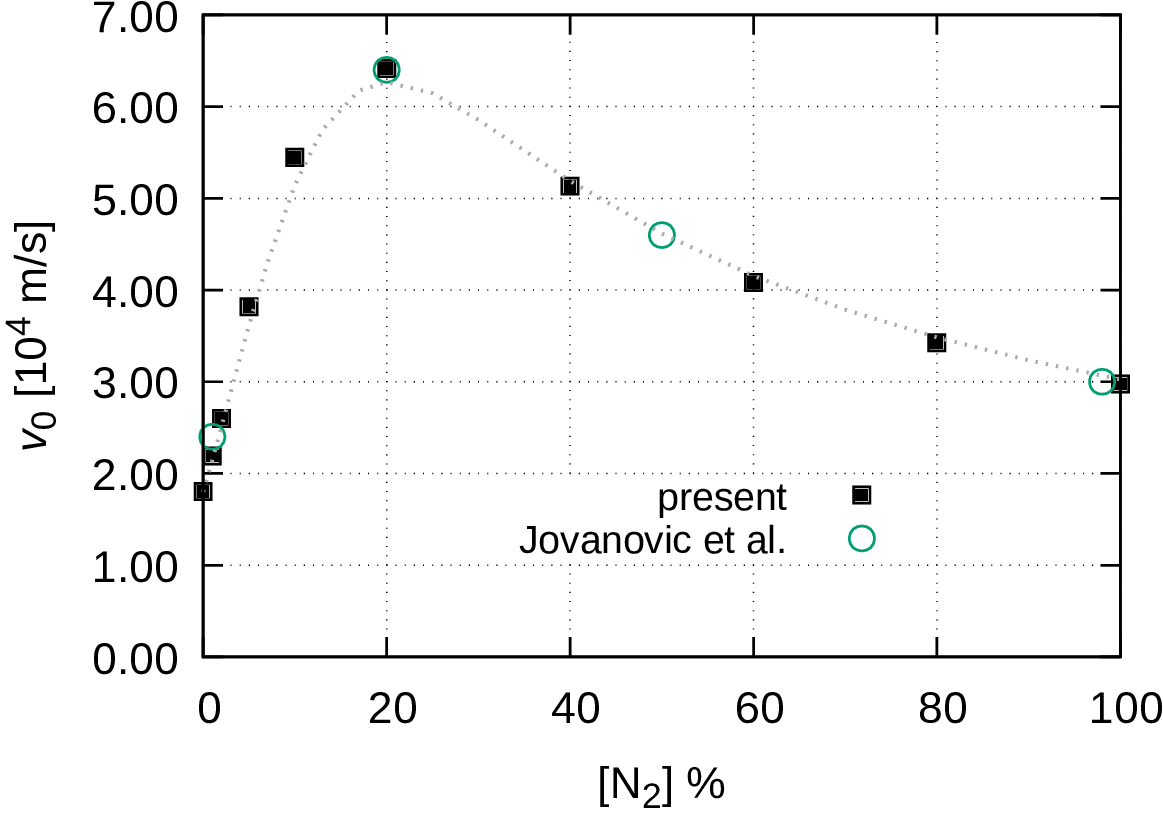}
\footnotesize(d) \includegraphics[width=0.45\textwidth]{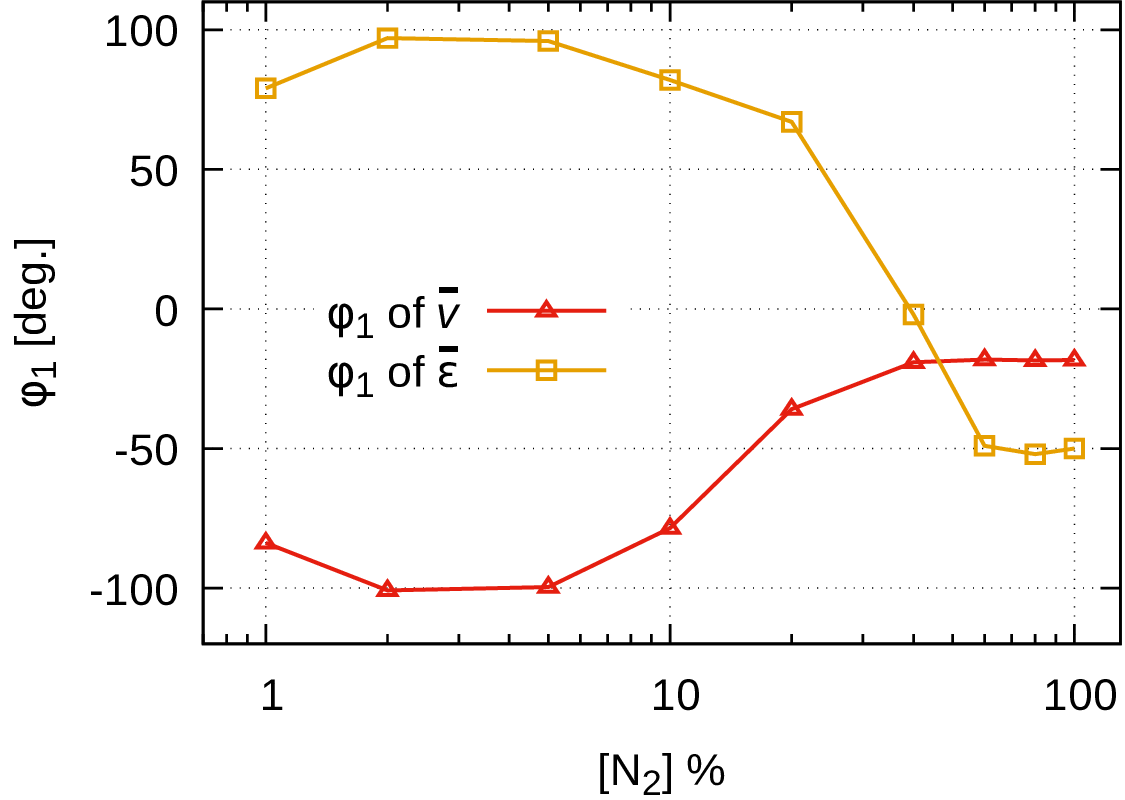}
\caption{Spatial profiles of the mean electron velocity ($\overline{v}(x)$) (a) and mean electron energy ($\overline{\varepsilon}(x)$) (b) as a function of the N$_2$ concentration. The legend in (a) also holds for (b). The data in (a) are normalised by the equilibrium velocity of electrons ($v_0$) that varies as a function of the Ar/N$_2$ mixing ratio as shown in (c). The results of our calculations are cross checked with the data of  \cite{Jovanovic}. (d) Phase of the first harmonic of $\overline{v}(x)$ and $\overline{\varepsilon}(x)$ as a function of the N$_2$ content. $(E/N)_0$ = 20 Td, $M$ = 0.2, and $L$ = 2.7 cm. The dashed black lines in (a) and (b) show the spatial variation of $E/N$, this curve is given without units.}
\label{fig:nitrogen1}
\end{center}
\end{figure}

\begin{figure}[ht ]
\begin{center}
\footnotesize(a) \includegraphics[width =0.5\textwidth]{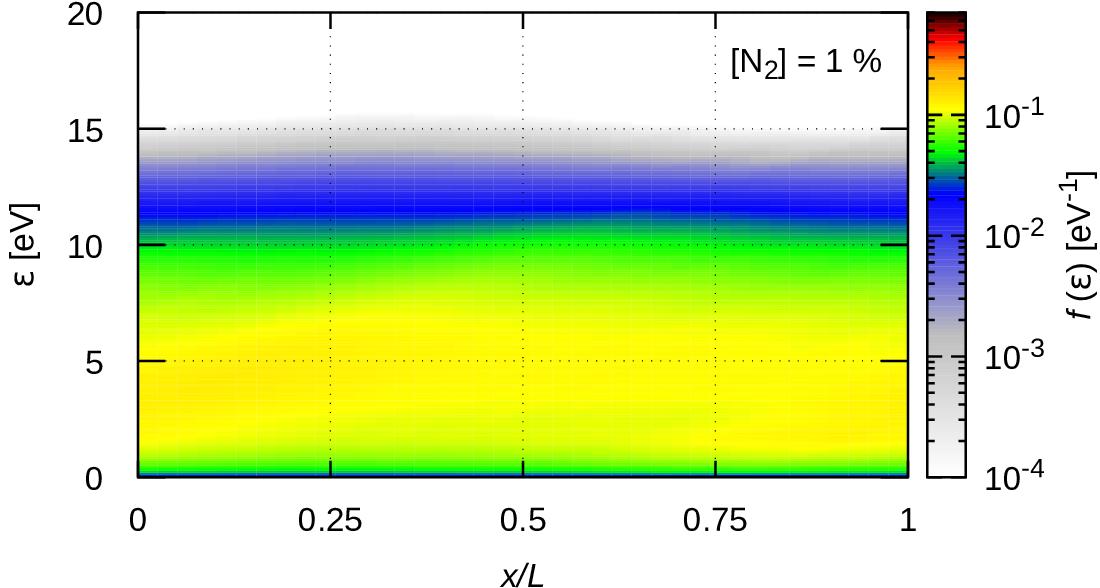}\\
\footnotesize(b) \includegraphics[width =0.5\textwidth]{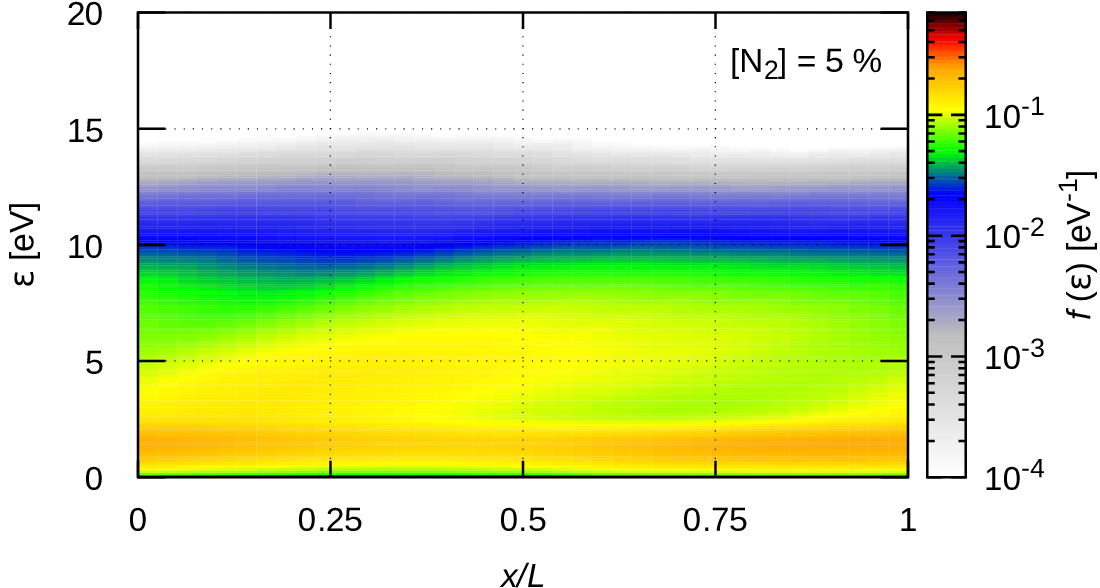}\\
\footnotesize(c) \includegraphics[width =0.5\textwidth]{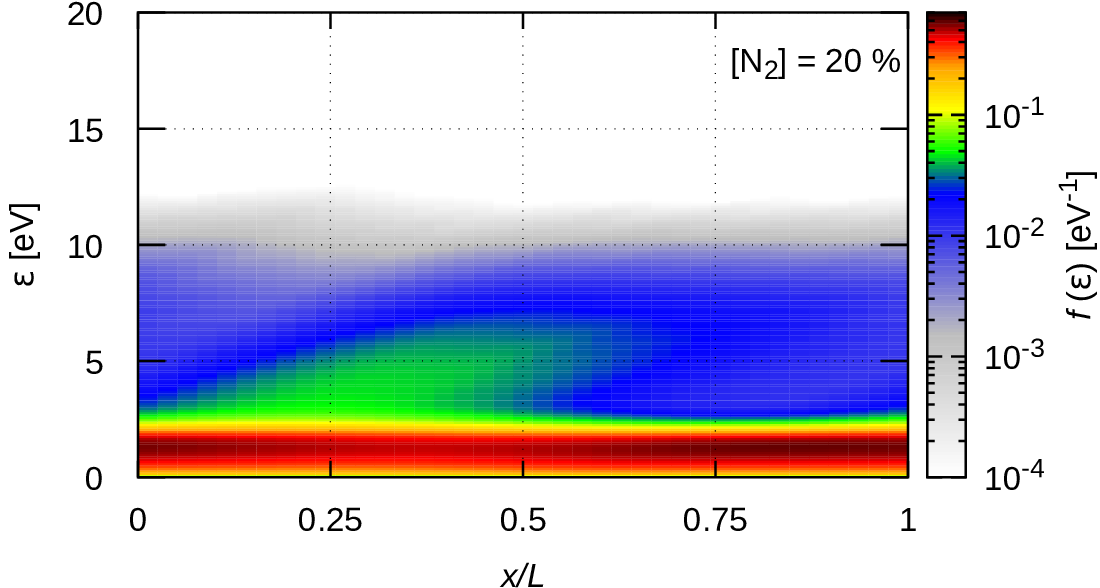}
\caption{Spatial maps of the EEDF for $(E/N)_0$ = 20 Td and $M$ = 0.2, and $L$ = 2.7 cm, for N$_2$ concentrations of (a) 1\,\%, (b) 5\,\%, and (c) 20\,\%.}
\label{fig:n2effect_edf}
\end{center}
\end{figure}

Figure \ref{fig:nitrogen1}(a) shows $\overline{v}(x)/v_0$ for various percentages of N$_2$ between 0\% and 100\%, for $(E/N)_0$ = 20 Td, $M$ = 0.2, and $L$\,=\,2.7 cm. The data are normalised by the equilibrium velocity $v_0$, which varies (as shown in figure \ref{fig:nitrogen1}(c)) as a function of the N$_2$ content in the gas mixture. (In the case of 0\% N$_2$ the data are the same as shown in figure \ref{fig:mod1}.) With an increasing N$_2$ percentage the phase of the first harmonic of $\overline{v}(x)$ first slightly decreases at low nitrogen content, and then increases rapidly to $\varphi_1 \approx -20^\circ$ above 5\% N$_2$ (see figure \ref{fig:nitrogen1}(d)). The velocity profiles practically overlap at $\geq$ 40\% N$_2$ content indicating that excitation of Ar is strongly suppressed at these molecular gas concentrations. Some of the data points belong to the parameter range where Negative Differential Conductivity in the Ar-N$_2$ mixture is present. As revealed from figure 4 of \cite{Dyatko}, NDC at $E/N$ = 20 Td occurs between N$_2$ concentrations of approx. 5\% and 15\%. This effect may have an influence on the behaviour of $\varphi_1$, clarification of this is, however, left for future work that needs to consider a broader domain of the parameters especially the spatial wavelength of the modulation. The addition of N$_2$ efficiently cools the electrons, as the profiles of the mean energy, shown in figure \ref{fig:nitrogen1}(b) confirm. Besides the value of $\varepsilon$ the  modulation of its spatial profile decreases as well and the phase of the profile changes also significantly as shown in figure \ref{fig:nitrogen1}(d).

The presence of N$_2$ in the mixture has a dramatic effects on the EEDF as well, as shown in figure \ref{fig:n2effect_edf}. Already at 1\% N$_2$ content, the marked spatial modulation of the EEDF observed in pure Ar (cf. figure \ref{fig:EEDF2}(b)) vanishes almost completely. With the addition of more N$_2$, the low energy part of the EDDF gets gradually more populated as a result of the low-threshold-energy processes in N$_2$. 

\begin{figure}[ht ]
\begin{center}
\footnotesize(a) \includegraphics[width =0.41\textwidth]{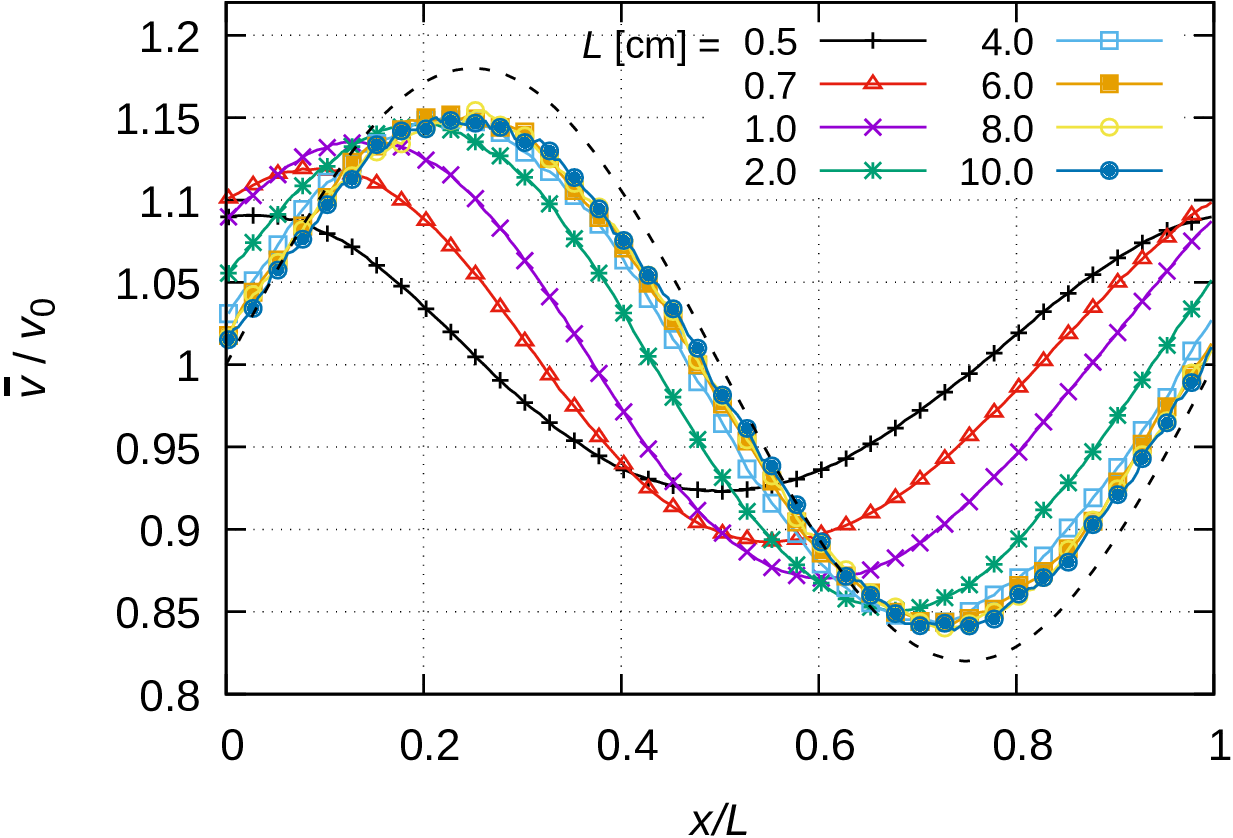}
\footnotesize(b) \includegraphics[width =0.41\textwidth]{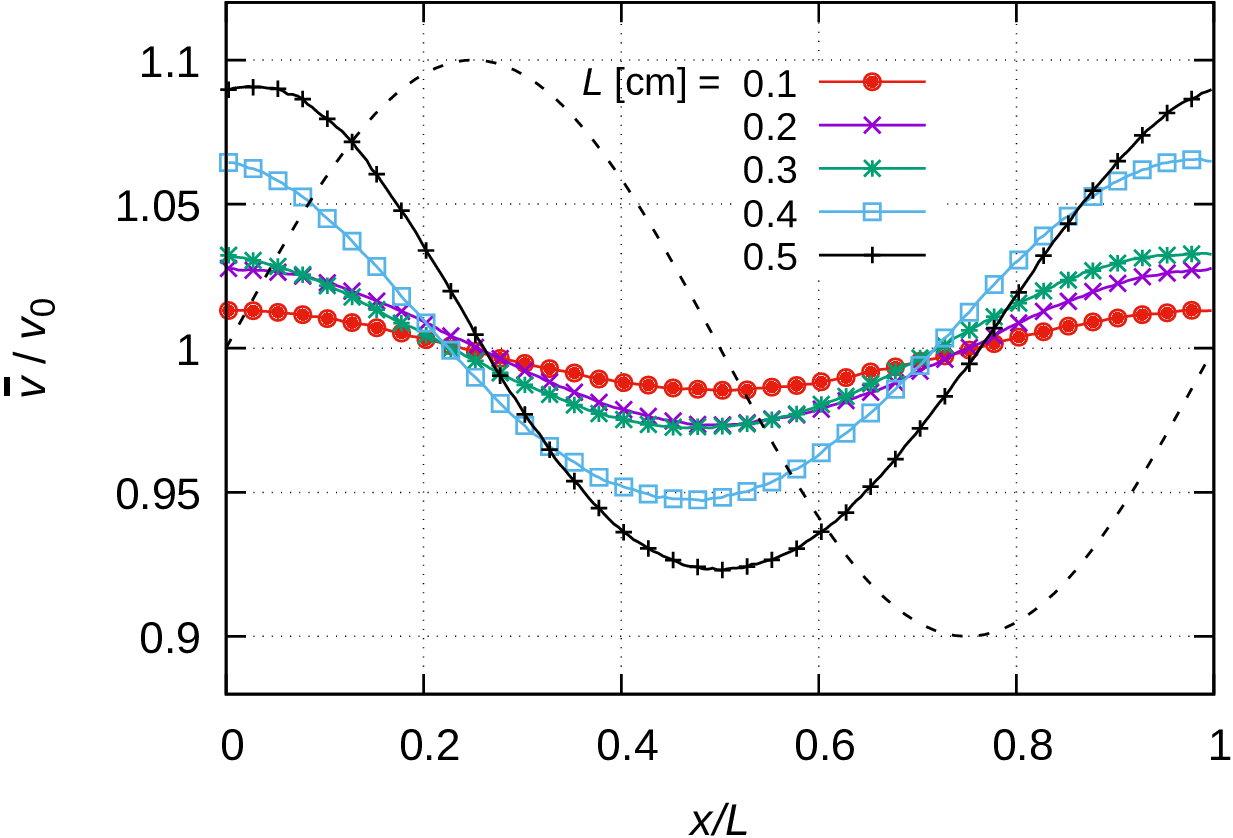}\\
\footnotesize(c) \includegraphics[width =0.41\textwidth]{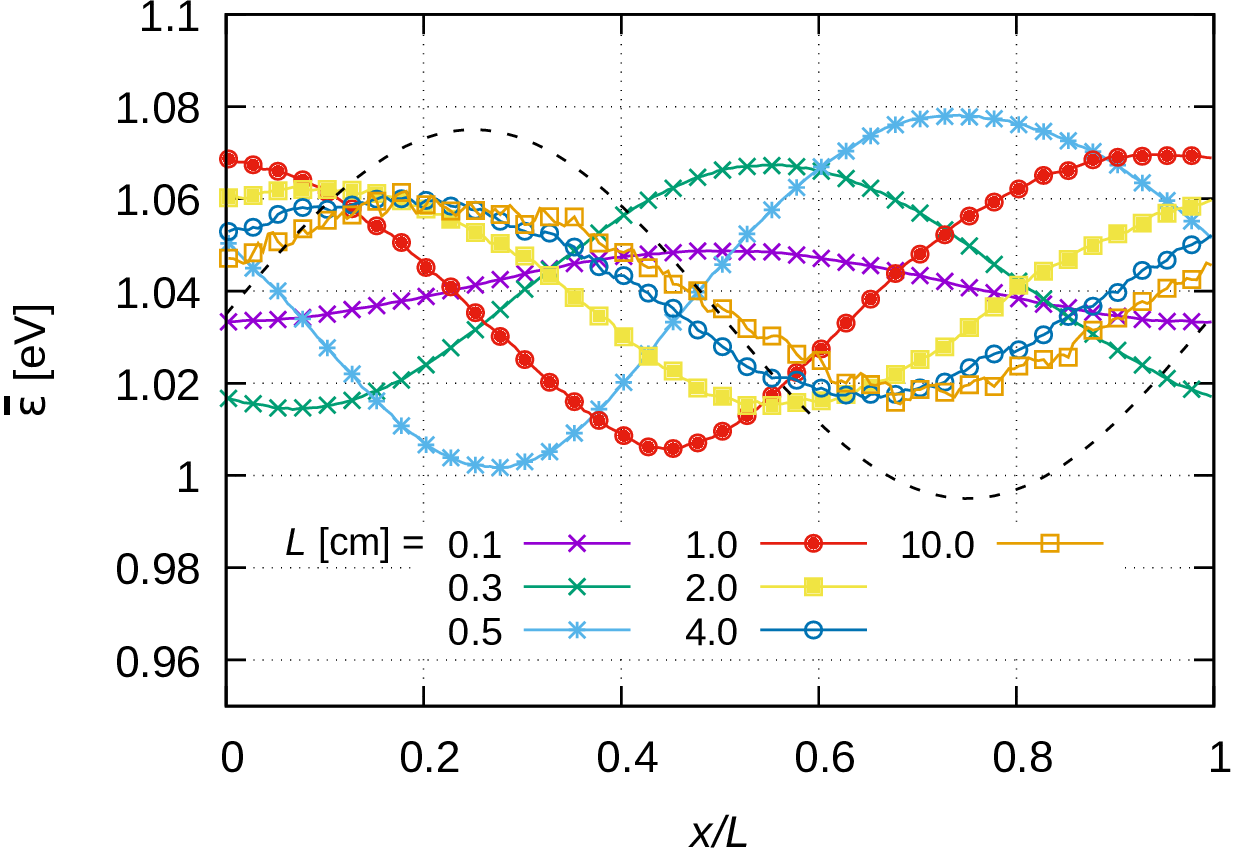}
\footnotesize(d)~~ \includegraphics[width=0.4\textwidth]{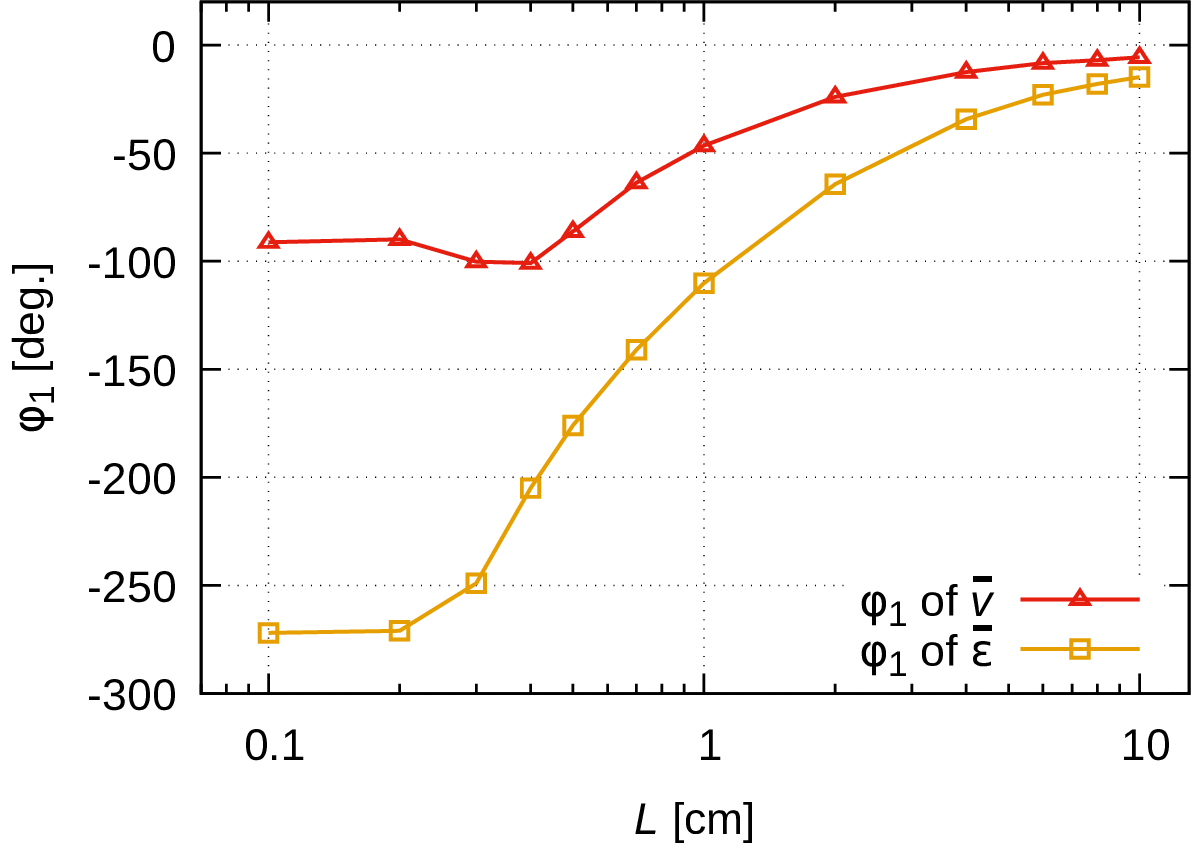}
\caption{(a,b) Normalized mean electron velocity in pure N$_2$ at $(E/N)_0$ = 20\,Td and $M$ = 0.2, for various values of $L$, given in the legends. (c) Profiles of the mean electron energy for the same conditions. The dashed black lines in each panel show the spatial variation of $E/N$, this curve is given without units. (d) Phase of first harmonic of $\overline{v}_{1}$ and $\overline{\varepsilon}_{1}$.}
\label{fig:nitrogen2}
\end{center}
\end{figure}

\begin{figure}[ht ]
\begin{center}
\footnotesize(a) \includegraphics[width =0.45\textwidth]{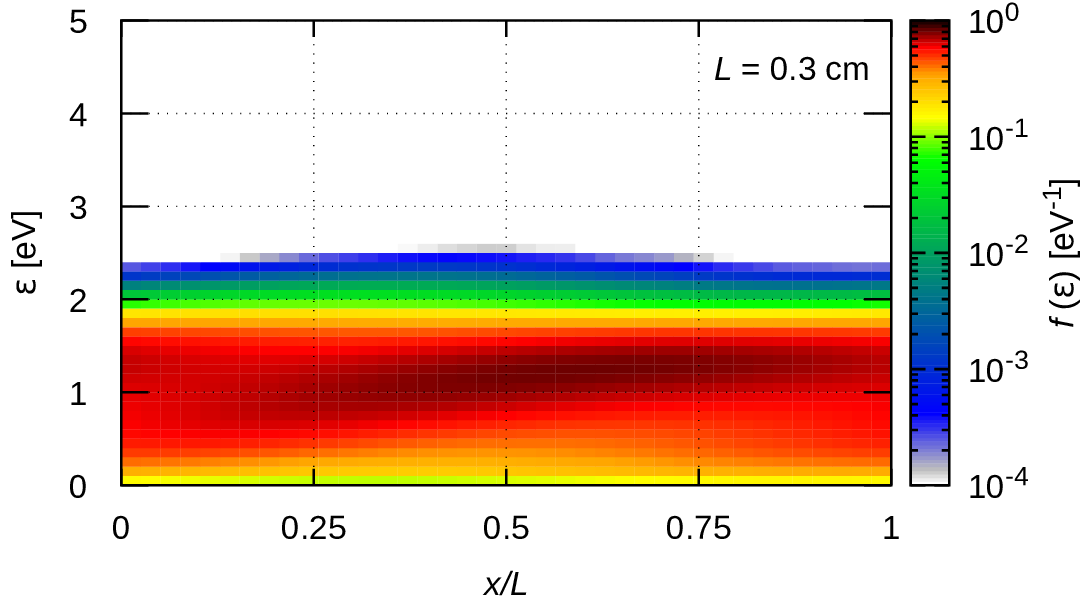}
\footnotesize(b) \includegraphics[width =0.45\textwidth]{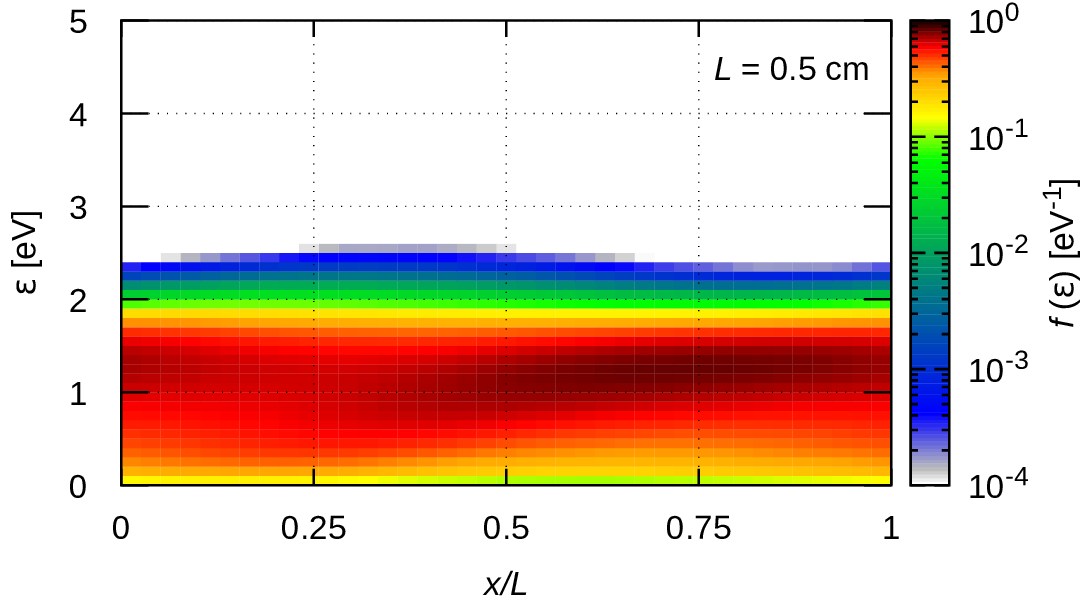}\\
\footnotesize(c) \includegraphics[width =0.45\textwidth]{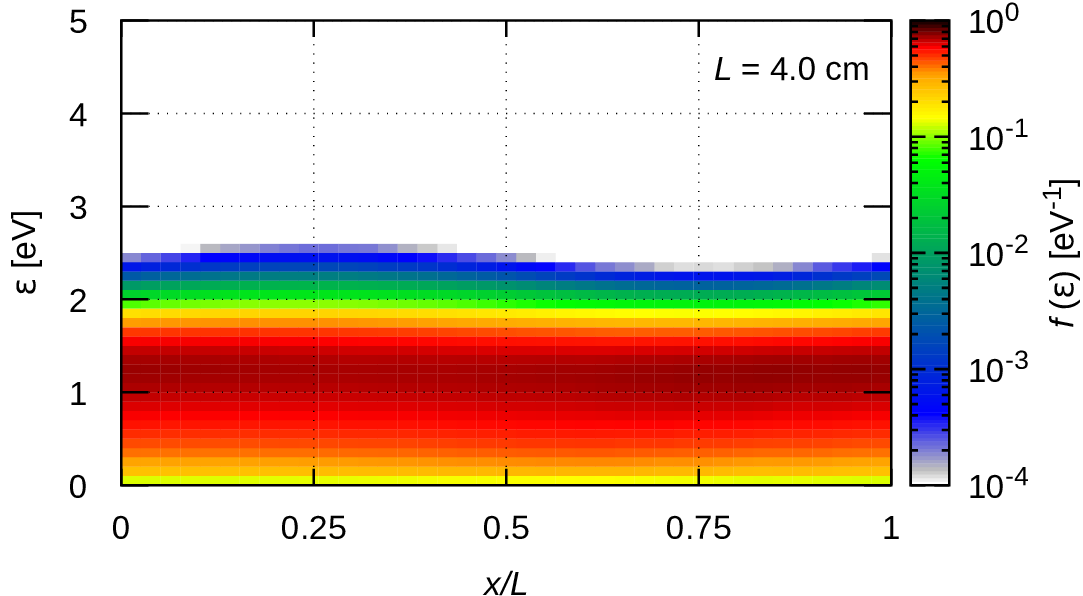}
\caption{Spatial maps of the EEDF in pure N$_2$, for $(E/N)_0$ = 20 Td and $M$ = 0.2, and various values of $L$.}
\label{fig:pure_n2_edf}
\end{center}
\end{figure}

Finally we present results for the case of pure N$_2$. Figures \ref{fig:nitrogen2}(a,b) display spatial profiles of the mean electron velocity for as a function of $L$, while panel (c) of the same figure shows the behaviour of the mean electron energy. At large $L$ (i.e. at $L >$2\,cm), both the $\overline{v}(x)$ and $\overline{\varepsilon}(x)$ profiles approximate the spatial dependence of the electric field. For these conditions, the phases of the first harmonic of both of these profiles approaches zero, as it is revealed from figure \ref{fig:nitrogen2}(d). These are signatures of the local character of the transport. For large $L$, we indeed find a very slight spatial modulation of the EEDF as well, as it can be seen in figure \ref{fig:pure_n2_edf}(c) for $L$ = 4\,cm. The only observable signature there is a small modulation of the high energy cutoff with $x/L$, around $\varepsilon \approx$ 2.5\,eV. The vast majority of the electrons have energies less than 2\,eV. For such energies, as figure \ref{fig:LF}(b) reveals, the energy relaxation length is in the order of $\lambda_{\rm e} \approx$ 1-2\,cm. For any $L$ exceeding this value we expect that the swarm properties reflect the local value of the electric field, as it is actually confirmed in figure \ref{fig:nitrogen2}(a).

\section{Summary}

\label{sec:summary}

In this work, we have investigated via Monte Carlo simulation the characteristics of electron transport in a stationary,  spatially modulated electric field. The computations have been executed for Ar and N$_2$ gases and their mixtures at (spatially averaged) reduced electric fields in the 10-40 Td range. Particles have been traced in a finite spatial region subjected to periodic boundary conditions.

Within the range of the reduced electric field considered, a strong response of the transport parameters to the electric field modulation was observed. At low modulation depths, the spatial profiles of the mean electron velocity and energy exhibited a harmonic shape. The phase angle between the electric field and the above quantities, as well as the harmonic content of the latter at higher modulation depths were revealed by Fourier analysis. All the quantities analysed showed highly nonlocal transport, except for the case of pure N$_2$ at long modulation wavelength, where signatures of local electron transport were observed.

At conditions, where the high modulation depth of $E/N$ resulted in the appearance of a region where the direction of the electric field is reversed, we observed trapping of the electrons. The stationary case, where all electrons are supposed to be trapped was not reached due to the slow accumulation of the electrons. The latter was found to be caused by the fact that the modulation of the electric field favours inelastic collisions (that represent high energy loss) outside the domain of the reversed field. For a realistic description of any experimental setting the inclusion of either Coulomb or thermal collisions is necessary.

Understanding the transport characteristics requires the analysis of the spatial variation of the electron energy distribution function (EEDF). We have computed this function for various parameter combinations and have found pronounced structures at the resonance condition where the potential drop over the simulation box is $\approx 13$\,V in pure Ar. The strongest resonance was found for $(E/N)_0$\,=\,20\,Td. At lower reduced electric field, elastic collisions play a more prominent role in the electron energy balance, while at higher field more excitation channels are open due to the higher electron energy, and this adversely affects the synchronisation of the kinetics of individual electrons. The analysis of the rates of the inelastic collision processes at 20 Td showed that for these conditions only few levels of the Ar atom are preferentially excited.  

Even small amounts of N$_2$ admixtures were found to lead to the vanishing of the structures in the EEDF due to the wider range of excitation energies of the N$_2$ molecule. 

Further directions of the present study include (i) the clarification of the combined effects of the modulated electric field and the Negative Differential Conductivity on the transport characteristics in the case of Ar-N$_2$ mixtures, (ii) investigations of the transport properties in the presence of non-sinusoidal perturbation of the electric field strength, (iii) inclusion of Coulomb collisions for an accurate prediction of the properties of the trapped part of the electron population in regions with reversed electric field, and (iv) the introduction of the self-consistent computation of the electric field distribution from the perturbed densities of the electrons via the Poisson equation in order to predict the range of existence of standing striations.

\ack This work was supported by the Hungarian National Office for Research, Development and Innovation (NKFIH) via the grant 119357. DB and SD are supported by the Ministry of Education, Science and Technological Development of the Republic of Serbia and the Institute of Physics (Belgrade).


\section*{References}
\begin{thebibliography}{99}

\bibitem{Winkler2002} Winkler R, Loffhagen D, Sigeneger F 2002
{\it Applied Surface Science} {\bf 192} 50

\bibitem{KumarSR1980} Kumar K, Skullerud H R and Robson R E 1980 {\it Aust. J. Phys.} {\bf 33} 343

\bibitem{noneq1} Robson R E, White R D and Petrovi\'c Z Lj 2005 {\it Rev. Mod. Phys.} {\bf 77} 1303
                         
\bibitem{noneq2} Pitchford L C, Boeuf J-P, Segur P and Marode E 1990 in {\it Non-equilibrium Effects in Ion and Electron Transport}, ed. Gallagher J W (Plenum, New York)

\bibitem{noneq3} Tsendin L D 1995 {\it Plasma Sources Sci. Technol.} {\bf 4} 200

\bibitem{noneq4} Kudryavtsev A A, Morin A V and Tsendin L D 2008 {\it Tech. Phys.} {\bf 53} 1029 

\bibitem{equi} Malovi\'c G, Strini\'c A, \v{Z}ivanov A, Mari\'c D, Petrovi\'c Z 2003 {\it Plasma Sources Sci. Technol.} {\bf 12} S1

\bibitem{Donko2011} Donk\'o Z 2011 {\it Plasma Sources Sci. Technol.} {\bf 20} 024001

\bibitem{FH} Franck J, Hertz G 1914 {\it Verh. Deut. Phys. Ges.} {\bf 16} 457

\bibitem{RobsonLW2000} Robson R E, Li B and White R D 2000 {\it J. Phys. B: At. Mol. Opt. Phys.} {\bf 33} 507

\bibitem{Sigeneger2003} Sigeneger F, Winkler R, Robson R E 2003 {\it Contrib. Plasma Phys.} {\bf 43} 178

\bibitem{White2012} White R D, Robson R E, Nicoletopoulos P, Dujko S 2012 {\it Eur. Phys. J. D} {\bf 66} 117 

\bibitem{Robson2014} Robson R E, White R D,  Hildebrandt M 2014 {\it Eur. Phys. J. D} {\bf 68} 188 

\bibitem{Loffhagen2002} Loffhagen D, Winkler R, Donk\'o 2002 {\it Eur. Phys. J. Appl. Phys.} {\bf 18} 189 

\bibitem{Dujko2008} Dujko S, White R D, Petrovi\'c 2008 {\it J. Phys. D} {\bf 41} 245205

\bibitem{White2009} White R D, Robson R E, Dujko S, Nicoletopoulos P, Li B 2009 {\it J. Phys. D: Appl. Phys.} {\bf 42} 194001

\bibitem{equilibration2019} Donk\'o Z,  Hartmann P, Korolov I, Jeges V,  Bo{\v{s}}njakovi{\'{c}} V, Dujko S 2019 {\it Plasma Sources Sci. Technol.} {\bf 28} 095007

\bibitem{Golubovsky1998} Golubovsky Yu B, Porokhova I A, Behnke J, Nekutchaev V O 1998
{\it J. Phys. D: Appl. Phys.} {\bf 31} 2447

\bibitem{Golubovskii1999} Golubovskii Y B, Maiorov V A, Porokhova I A, Behnke J 1999
{\it Journal of Physics D: Applied Physics} {\bf 32} 1391

\bibitem{Kolobov2006} Kolobov V I 2006 
{\it Journal of Physics D: Applied Physics} {\bf 39} R487

\bibitem{Klarfeld1952} Klarfeld B N 1952
{\it Sov. Phys.-JETP} {\bf 22} 66

\bibitem{Pekarek1962} Pekarek L,  Krej\v{c}\'{i} 1962
{\it Cechoslovackij fiziceskij zurnal B} {\bf 12} 450

\bibitem{Sigeneger2000a} Sigeneger F, Sukhinin G I,  Winkler R 2000
{\it Plasma Chemistry and Plasma Processing} {\bf 20} 87

\bibitem{Sukhinin2006} Sukhinin G I, Fedoseev A V 2006
{\it High temperature} {\bf 44} 157

\bibitem{Raizer1997} Raizer Yu P, Shneider M N 1997
{\it High Temp.} {\bf 35} 19

\bibitem{Iza2005} Iza F, Yang S S, Kim H C, Lee J K 2005 
{\it Journal of Applied Physics} {\bf 98} 043302

\bibitem{Stittsworth1996} Stittsworth J A, Wendt A E 1996
{\it IEEE Transactions on Plasma Science} {\bf 24} 125

\bibitem{Sigeneger1998} Sigeneger F, Golubovskii Yu B,
Porokhova I A, Winkler R 1998 {\it Plasma Chemistry and Plasma Processing} {\bf 18} 153

\bibitem{Sigeneger2000b} Sigeneger F, Winkler R 2000 {\it Plasma Chemistry and Plasma Processing} {\bf 20} 429

\bibitem{Liu2016} Liu Y X, Schuengel E, Korolov I, Donk\'o Z, Wang Y N, Schulze J 2016
{\it Physical Review Letters} {\bf 116} 255002

\bibitem{Arslanbekov2019} Arslanbekov R A, Kolobov V I 2019
{\it Phys. Plasmas} {\bf 26} 104501 

\bibitem{Kolobov2020} Kolobov V I, Arslanbekov R A, Levko D, Godyak V A 2020 {\it J. Phys. D: Appl. Phys.} accepted manuscript JPhysD-123530.R1

\bibitem{Harti2020} Hartmann P, Rosenberg M, Juhasz Z, Matthews L S, Sanford D L, Vermillion K, Reyes J C, Hyde T W 2020 {\it Plasma Sources Science and Technology} http://iopscience.iop.org/10.1088/1361-6595/abb955

\bibitem{Golubovskii12013} Golubovskii Yu B, Kolobov V I, Nekuchaev V O 2013
{\it Physics of Plasmas} {\bf 20} 101602 

\bibitem{Denpoh} Denpoh K 2012 {\it Japanese Journal of Applied Physics} {\bf 51} 106202

\bibitem{Liu2019} Liu Y-X, Donk\'o Z, Korolov I, Schuengel E, Wang Y-N, Schulze J 2019 {\it Plasma Sources Science and Technology} {\bf 28}

\bibitem{Shvydky} Shvydky A A, Khudik V N, Nagorny V P, Theodosiou C E 2006 {\it IEEE Transactions on Plasma Science} {\bf 34} 878

\bibitem{Tagashira86} Sato N, Tagashira H 1985 {\it J. Phys. D: Appl. Phys.} {\bf 18} 2451

\bibitem{Boeuf82} Boeuf J P and Marode E 1982 {\it J. Phys. D: Appl. Phys.} {\bf 15} 2169

\bibitem{Hayashi} Hayashi M, Nagoya Institute of Technology Report IPPJ-AM- 19 (unpublished).

\bibitem{Biagi} Biagi database, www.lxcat.net, retrieved on July 23, 2020

\bibitem{NDC} Petrovi\'c Z Lj, Crompton R W, Haddad G N 1984 {\it Aust. J. Phys.} {\bf 37} 23

\bibitem{Dyatko} Dyatko N A, Kochetov I V, Napartovich A P 2014 {\it Plasma Sources Sci. Technol.} {\bf 23} 043001 

\bibitem{Nicoletopoulos} Nicoletopoulos P, Robson R E,  White R D 2012 {\it Phys. Rev. E} {\bf 85} 046404

\bibitem{Aleksandrov} Aleksandrov N L, Kochetov I V 1996 {\it J. Phys. D: Appl. Phys.} {\bf 29} 1476 

\bibitem{reversed} Loffhagen D, Sigeneger F, Winkler R 2004 {\it Eur. Phys. J. Appl. Phys.} {\bf 25} 45

\bibitem{FR1} Boeuf J P, Pitchford L C 1995 {\it J. Phys. D Appl. Phys.} {\bf 28} 2083

\bibitem{FR2} Kudryavtsev A A, Nisimov S U, Prokhorova E I, Slyshov A G 2011 {\it Technical Physics Letters} {\bf 37} 838

\bibitem{Kushner} Weng Y, Kushner M J 1990 {\it Phys. Rev. A} {\bf 42} 6192 

\bibitem{Jovanovic} Jovanovi\'c J V, Basurto E, \v{S}a\v{s}i\'c O, Hern\'andez-\'Avila J L, Petrovi\'c Z L, De Urquijo J 2009 {\it Journal of Physics D: Applied Physics} {\bf 42} 045202


\end{thebibliography}
\end{document}